\documentclass[11pt,a4paper]{article}
\usepackage{jinstpub} 
\usepackage{graphicx} 
\usepackage{xcolor}

\newcommand{\bsmcitations}{\cite{sterile,sterile2,supersymmetry,sphaleron,cptsymmetric,stau,leptoquark,rparity,darkmatter,shdm,axions,prosandcons}}

\title{The Payload for Ultrahigh Energy Observations (PUEO):\\ A White Paper}
\abstract
{
The Payload for Ultrahigh Energy Observations (PUEO) long-duration balloon
  experiment is designed to have world-leading sensitivity to ultrahigh-energy
  neutrinos at energies above 1 EeV. Probing this energy region is essential
  for understanding the extreme-energy universe at all distance scales. PUEO
  leverages experience from and supersedes the successful Antarctic Impulsive
  Transient Antenna (ANITA) program, with an improved design that drastically
  improves sensitivity by more than an order of magnitude at energies below 30 EeV.  PUEO will either make the first significant detection of
  or set the best limits on ultrahigh-energy neutrino fluxes. 
}

\author[a]{Q.~Abarr}
\author[b]{P.~Allison}
\author[c]{J.~Ammerman~Yebra}
\author[c]{J.~Alvarez-Mu\~niz}
\author[b]{J.~J.~Beatty}
\author[d,e]{D.~Z.~Besson}
\author[f]{P.~Chen}
\author[f]{Y.~Chen}
\author[g]{J.~M.~Clem}
\author[b]{A.~Connolly}
\author[h]{L.~Cremonesi}
\author[i]{C.~Deaconu}
\author[b]{J.~Flaherty}
\author[b]{D.~Frikken}
\author[j]{P.~W.~Gorham}
\author[k]{C.~Hast}
\author[d]{C.~Hornhuber}
\author[l]{J.~J.~Huang}
\author[i]{K.~Hughes}
\author[m]{A.~Hynous}
\author[n]{Y.~Ku}
\author[f]{C.-Y.~Kuo}
\author[l]{T.~C.~Liu}
\author[j]{Z.~Martin}
\author[j]{C.~Miki}
\author[f]{J.~Nam}
\author[o]{R.~J.~Nichol}
\author[j]{K.~Nishimura}
\author[d]{A.~Novikov}
\author[d]{A.~Nozdrina}
\author[i]{E.~Oberla}
\author[b]{S.~Prohira}
\author[j]{R.~Prechelt}
\author[a]{B.~F.~Rauch}
\author[p]{J.~M.~Roberts}
\author[q]{A.~Romero-Wolf}
\author[j]{J.~W.~Russell}
\author[g]{D.~Seckel}
\author[f]{J.~Shiao}
\author[i]{D.~Smith}
\author[i]{D.~Southall}
\author[j]{G.~S.~Varner}
\author[i]{A.~G.~Vieregg}
\author[f]{S.-H.~Wang}
\author[f]{Y.-H.~Wang}
\author[n,r,s]{S.~A.~Wissel}
\author[o]{C.~Xie}
\author[d]{R.~Young}
\author[c]{E.~Zas}
\author[n]{A.~Zeolla}

\affiliation[a]{Dept. of Physics, McDonnell Center for the Space Sciences, Washington Univ. in St. Louis, MO 63130.}
\affiliation[b]{Dept. of Physics, Center for Cosmology and AstroParticle Physics, Ohio State Univ., Columbus, OH 43210.}
\affiliation[c]{Instituto Galego de F\'isica de Altas Enerx\'ias (IGFAE), Universidade de Santiago de Compostela, 15782 Santiago de Compostela, Spain}
\affiliation[d]{Dept. of Physics and Astronomy, Univ. of Kansas, Lawrence, KS 66045.}
\affiliation[e]{Moscow Engineering Physics Institute, Moscow, Russia.}
\affiliation[f]{Dept. of Physics, Grad. Inst. of Astrophys., Leung Center for Cosmology and Particle Astrophysics, National Taiwan University, Taipei, Taiwan.}
\affiliation[g]{Dept. of Physics, Univ. of Delaware, Newark, DE 19716.}
\affiliation[h]{School. of Physics and Astronomy, Queen Mary University of London, London, United Kingdom.}
\affiliation[i]{Dept. of Physics, Enrico Fermi Inst., Kavli Inst. for Cosmological Physics, Univ. of Chicago, Chicago, IL 60637.}
\affiliation[j]{Dept. of Physics and Astronomy, Univ. of Hawaii, Manoa, HI 96822.}
\affiliation[k]{SLAC National Accelerator Laboratory, Menlo Park, CA, 94025.}
\affiliation[l]{Dept. of Physics,  National Yang Ming Chiao Tung University, Taipei, Taiwan.}
\affiliation[m]{NASA Wallops Flight Facility, Wallops Island, VA, 23337}
\affiliation[n]{Pennsylvania State Physics Dept., Inst. for Gravitation and the Cosmos, University Park, PA 16802.}
\affiliation[o]{Dept. of Physics and Astronomy, University College London, London, United Kingdom.}
\affiliation[p]{Center for Astrophysics and Space Sciences, Univ. of California, San Diego, La Jolla, CA 92093.}
\affiliation[q]{Jet Propulsion Laboratory, California Institute for Technology,  Pasadena, CA 91109.}
\affiliation[r]{Pennsylvania State Astronomy and Astrophysics Dept., University Park, PA 16802.}
\affiliation[s]{California Polytechnic State Univ., Physics Dept., San Luis Obispo, CA 93407}

\collaboration{PUEO Collaboration}

 \emailAdd{cozzyd@kicp.uchicago.edu}
\emailAdd{avieregg@kicp.uchicago.edu}

\begin{document}
\maketitle

\section{Introduction}

\noindent Ultrahigh-energy (UHE) neutrinos ($\gtrsim$ 1 EeV) have long been
predicted but so far have eluded detection. The Payload for Ultrahigh Energy
Observations (PUEO) is a long-duration balloon experiment that will have
world-leading sensitivity to broad swaths of unexplored parameter space in UHE
neutrino flux using the radio-detection technique from a high-altitude
platform.  PUEO builds on the foundation laid by the Antarctic Impulse
Transient Antenna (ANITA) program, which is the best existing probe of UHE flux
above 30 EeV.  The improved sensitivity of PUEO is a result of a number of
novel features compared to ANITA, including a larger antenna array and a
beam-forming trigger. PUEO will be able to measure or rule out a number of
viable UHE models, providing better understanding of the UHE accelerators
throughout the entire universe. 

In this white paper, we 
lay out the science case for UHE neutrinos and PUEO,
review the radio-detection technique,
summarize the results from the ANITA program, introduce the PUEO design, and
justify PUEO's expected science performance.

\section{The PUEO Science Program}

High-energy neutrino astrophysics evinces a unique perspective on the energetic
particles from cosmic distances. Unlike other high-energy particle messengers, neutrinos
are unimpeded as they make their way across the universe, carrying information
about distant sources that is not accessible otherwise. Detection of UHE
neutrinos is the cornerstone of the PUEO science program. 
UHE neutrinos may be
produced 
in several different ways, further described below. Additionally, PUEO
is able to measure air showers from energetic particles in the atmosphere and
study Antarctic ice.

\subsection{Cosmogenic Neutrinos} 
\begin{figure}

  \centering
\includegraphics[width=0.49\textwidth]{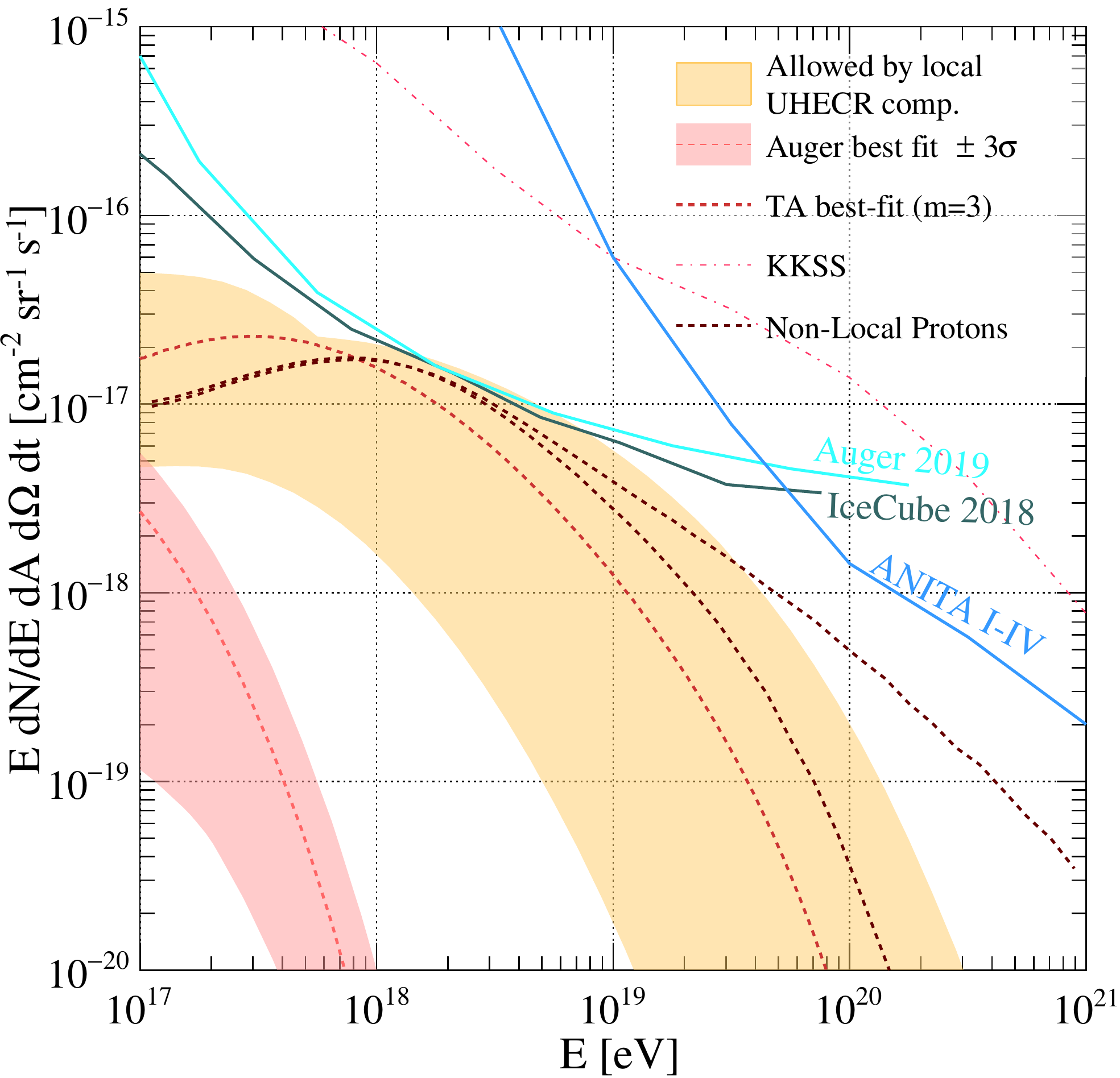}
\includegraphics[width=0.49\textwidth]{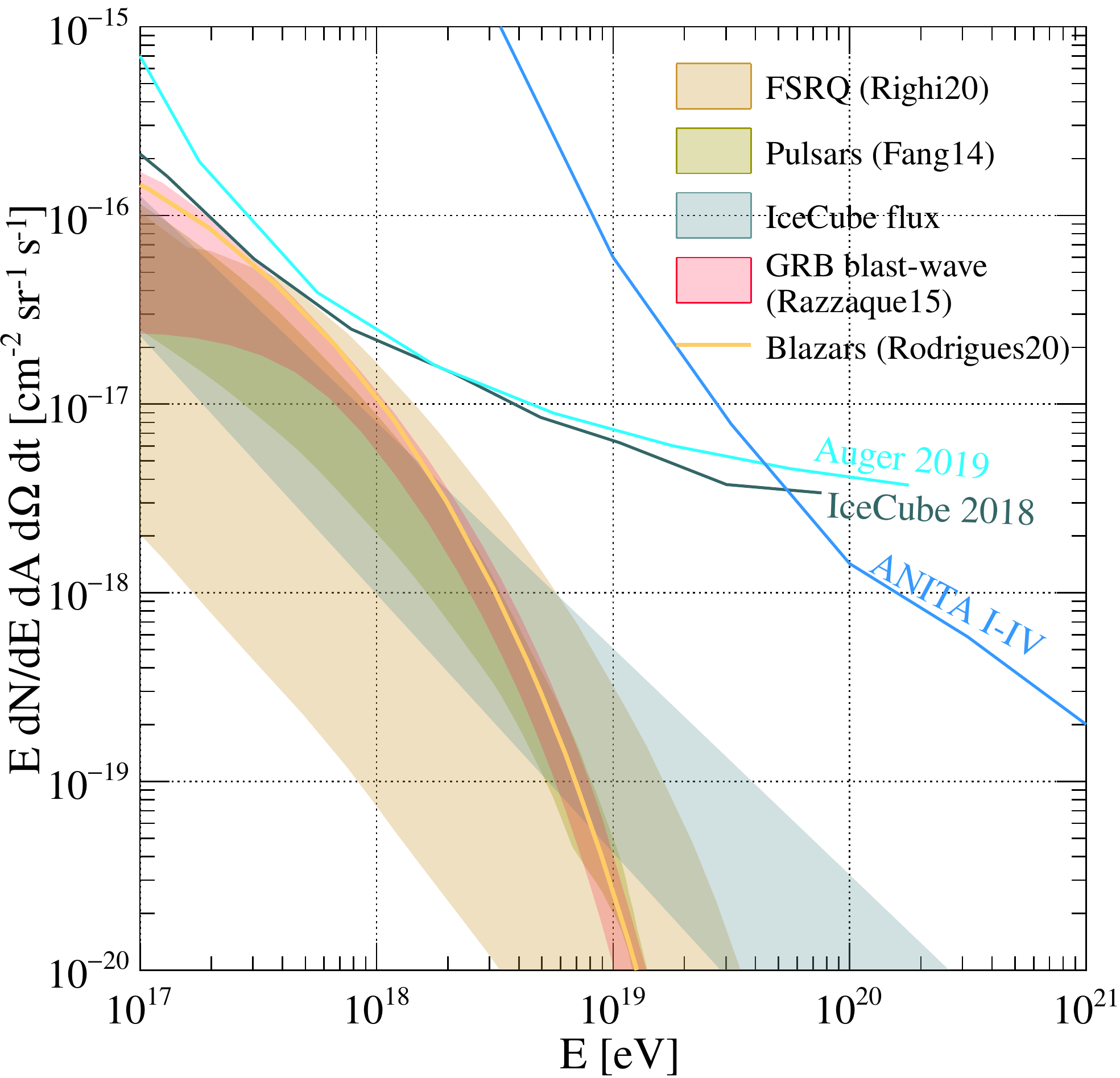}
  \caption{ \textbf{\textit{Left:}} Current limits~\cite{Aartsen:2018vtx,auger2019,anita4} on Cosmogenic Neutrinos and some viable models. The models displayed are uniform source class best fits to PAO data~\cite{Heinze:2019jou} (with uncertainties), similar fits to TA data~\cite{TA_composition_1} with a fixed source evolution model, subdominant all-proton models allowed by the measured proton fraction at PAO and TA~\cite{vanVliet:2019nse, TA_composition}. As explained in the text, the non-local proton source models were developed with CRPropa3~\cite{Batista:2016yrx} using a similar procedure as the subdominant models in ~\cite{vanVliet:2019nse}, but assuming sources have $z>0.1$. Additionally, the KKSS~\cite{kkss} model is shown as an example of a cosmogenic model that has been ruled out by ANITA. \textbf{\textit{Right:}} The same experimental limits but with diffuse astrophysical flux models, including FSRQs~\cite{righi2020eev}, AGN~\cite{rodrigues2020blazar}, GRB blast-waves~\cite{GRBBlastwave}, pulsars~\cite{FangPulsar} and an extrapolation of the IceCube flux~\cite{IceCubeFlux}.}
\label{fig:current_limits} 
\end{figure}

Cosmogenic neutrinos are produced from interactions of the UHECRs with the
cosmic-microwave background as they propagate through the universe. For
protons, this is called the Greisen-Zatsepin-Kuzmin (GZK) process
\cite{Greisen, ZatsepinK} and has a threshold of 50 EeV and a typical
length-scale of 50-200 Mpc, producing neutrinos on average with 5\% of the
proton energy~\cite{BerezinskyZ}. Higher-mass UHECRs may also produce neutrinos,
albeit less efficiently, through photodisintegration~\cite{Stecker:1998ib}.

UHECRs have been measured up to hundreds of EeV~\cite{Deligny:2020gzq}, but
the sources are still unknown and the chemical composition at Earth is still uncertain, leading to
significant uncertainties in predicted cosmogenic neutrino fluxes, even if we assume that the local universe is a representative sample of the entire universe. Conversely, UHE neutrino flux
measurements uniquely probe cosmic ray acceleration and mass
composition~\cite{seckel,vanVliet:2019nse} at all distance scales and complement cosmic rays in source
identification~\cite{Beresinsky:1969qj,Beresinsky:1975,Stecker:1978ah,Hill:1983xs,Yoshida:1993pt,Engel:2001hd,Anchordoqui:2007fi,takami,Ahlers:2009rf,Ahlers:2010fw,kotera,Yoshida:2012gf,Ahlers:2012rz,Aloisio:2015ega,Heinze:2015hhp,Romero-Wolf:2017xqe,AlvesBatista:2018zui}.
Existing experiments have already ruled out some cosmogenic
models~\cite{anita1,kkss,IceCube2016, Aartsen:2018vtx}.  PUEO's improved sensitivity will
either detect cosmogenic neutrinos or rule out 
scenarios where the UHECRs contain a sizable proton
component, their sources reach extreme 
maximum acceleration energies, and/or are more
populous at large redshifts~\cite{kotera,kkss}.
Fig.~\ref{fig:current_limits} (left) shows existing constraints on UHE neutrino
fluxes, as well as several still-viable models of cosmogenics fluxes (and one model that has been ruled out).

Cosmogenic neutrino fluxes from sources within the GZK horizon may be
constrained by current UHECR composition measurements. Measurements of the
penetration depth of energetic air showers in the Pierre Auger Observatory
(PAO) currently favor a mixed composition at Earth, although a proton
subcomponent is still allowed~\cite{auger16,Auger17}. The Telescope Array's (TA)
analogous measurements report compatibility with a lighter mass
composition~\cite{TA,Abbasi_2018}. The composition and spectra from TA or PAO
can be used to fit a parametric model of the source properties using
cosmological propagation codes such as CRPropa~\cite{Batista:2016yrx}, which
allow estimation of the resulting  UHECR and cosmogenic particles at Earth given assumptions about source distribution and injection properties.
Depending on if the TA or PAO composition is used, very different cosmogenic neutrino fluxes may be predicted~\cite{AlvesBatista:2018zui,TA_composition_1,Romero-Wolf:2017xqe}.  It is also possible that the discrepancies are due to the different views of the sky between TA and PAO. Such a situation could arise if there is a light UHECR source within the GZK horizon in the Northern Hemisphere but not the Southern.

Fits for cosmogenic neutrino fluxes based on UHECR spectra and composition
usually rely on simplified parametric models, and, moreover, assume that all
UHECR sources are of the same type (even if the cosmological evolution is varied). 
If the unjustified assumption that all
nuclear species come from the same source classes is relaxed, models that produce more
copious amounts of cosmogenic neutrinos become viable~\cite{vanVliet:2019nse,
TA_composition}, as neither PAO nor TA data can exclude a subdominant
all-proton component to UHECR at the highest energies.  Moreover, significant
systematic uncertainties from uncertain hadronic physics are present in the
estimation of the UHECR mass composition, with different choices of hadronic
models producing significantly different local composition
estimates~\cite{AlvesBatista:2019tlv}. Currently, none of the hadronic models
used in modeling of depth of shower maximum can predict all aspects of
measurements~\cite{PhysRevLett.117.192001}. Detection or further limits on
cosmogenic fluxes can constrain UHECR mass composition independent of the complex
hadronic physics.

\begin{figure} 
  \centering
  \includegraphics[width=5in]{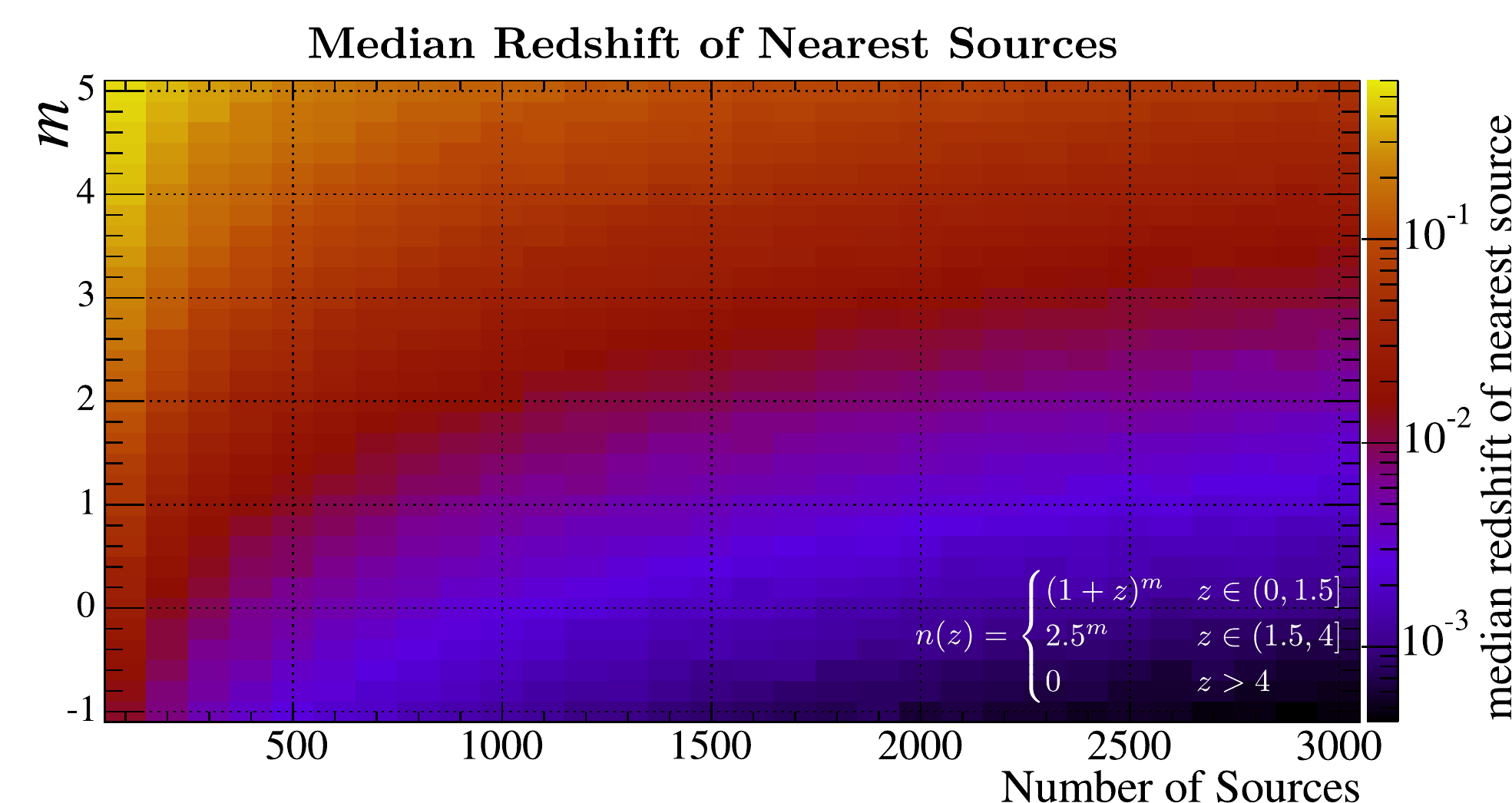} 

  \caption{The median distance of the nearest UHECR accelerator as a function
  of number of prototypes and the source evolution parameter $m$, where
  redshift is parametrized as $n(z) = (1+z)^{m}$ if $z \in (0,1.5]$, $n(z) =
  2.5^m$ if $z \in (1.5,4]$ and $n(z)=0$ if $z> 4$. This is calculated by
  drawing the number of sources from the source evolution 1000 times and
  computing the median of that distribution. This is generally a right-skewed
  distribution, so realizations where the nearest red shift is two or three
  times the median are not uncommon. }

  \label{fig:closest} 
\end{figure}

Published cosmogenic neutrino models based on current UHECR measurements do not
fully account for the possibility that some UHECR source classes may fluctuate to have no
prototypes within the GZK horizon, a plausible scenario for source classes with
few objects and/or strongly positive source evolution.  As an example, there are no known
FSRQs with a redshift smaller than 0.1~\cite{2012ApJ...751..108A}.
More generally, in Fig.~\ref{fig:closest} we show the median redshift of the closest source as a function of number of sources and strength of source evolution. The yellow-to-red areas in that plot denote parameter space where the nearest accelerator is plausibly  at or beyond the GZK horizon. 

To study resulting fluxes from such source classes, we have run some
proton-only CRPropa simulations simular to those in~\cite{vanVliet:2019nse},
where we have enforced a minimum red shift of 0.1, therefore highly suppressing
the protons that reach Earth without suppressing the neutrinos. Some example
models that result in a subdominant UHECR proton flux compatible with all
composition measurements
and also not violating
current neutrino constraints are shown in Fig.~\ref{fig:current_limits} as
``non-local protons." Only UHE neutrino detectors such as PUEO have the ability
to probe such accelerators across the entirety of the universe. 

As essentially all protons significantly above the GZK
cutoff interact prior to reaching Earth, UHECR measurements set nearly no
constraints on the maximum energy ($E_{max}$) of proton sources~\cite{seckel},
but this information is imprinted in the spectrum of the cosmogenic neutrinos. The full picture of UHECR accelerators can only be completely revealed with the aid of UHE neutrinos.

\subsection{UHE Astrophysical Neutrinos} 

UHE neutrinos may also be produced directly in astrophysical sources, through
hadronic or photohadronic processes, rather than in the propagation of the
UHECRs. These astrophysical neutrinos have been measured up to multiple PeV by
IceCube~\cite{IceCube2013a, IceCube2014a, IceCube2015a, IceCubeFlux}. It is
possible that this, or another, astrophysical flux may extend to energies that PUEO can probe. 

Astrophysical objects that could produce neutrinos at EeV energies include
GRBs, pulsars, magnetars from neutron-star mergers, and
FSRQs~\cite{righi2020eev, rodrigues2020blazar,FangPulsar, grb_fireball,
Metzger,Murase:2007yt, GRBBlastwave}.  In general, UHE neutrinos may be produced in sources
without saturating UHECR or gamma-ray bounds, so a UHE neutrino detector like
PUEO is required to explore this parameter space.  Depending on the number of
sources and their distances, the flux from astrophysical neutrinos may appear
diffuse or may be resolvable with stacking searches.
Fig.~\ref{fig:current_limits} (right) shows some of the astrophysical models
that produce the highest-energy neutrinos. 

Of particular interest to PUEO are transient astrophysical sources of UHE
neutrinos that produce large fluences in a short time window (e.g. GRBs,
Supernovae, FRBs, flaring blazars, neutron-star mergers). For this class of
sources, PUEO is an ideal instrument due to its world-leading instantaneous
aperture at UHE energies within its field of view, which allows for detecting
fluences above background levels. For transients with multimessenger
measurements, the analysis efficiency may be increased and backgrounds
substantially reduced by focusing in on a narrow region of time and
space~\cite{anitaGRB, anita3source}, further improving the sensitivity for UHE
detection.  As seen in the apparent IceCube event coincident with a flare TXS
0506+056~\cite{IceCube:2018cha,IceCube:2018dnn}, multimessenger coincidences
have the capability to more confidently attribute neutrinos to sources.

In addition to after-the-fact multimessenger correlation searches, in some
cases it may be beneficial for observatories to respond to prompt alerts for
followup~\cite{amon}. Such capabilities are not currently part of the PUEO
design, but once sufficient experience is established successfully
operating PUEO and understanding its anthropogenic noise environment, sending
out alerts for high-probability events over a prioritized satellite stream could
eventually become possible in subsequent flights.

As will be further reviewed in Sec.~\ref{sec:anita}, ANITA-III has observed
candidate events that occur in spatial and temporal coincidence with
supernovae~\cite{anita3me,anita3source}(SN). While these ANITA associations are not statistically-significant, PUEO's increased sensitivity will confidently
exclude or detect such a flux.

\subsection{Dark Matter and Top-Down Models} 

UHE neutrinos may also be produced directly, rather than in accelerators,  for
example in the decays of heavy dark matter (DM)~\cite{DM} or other top-down
models~\cite{td}. ANITA has already ruled out many such models, for instance
Z-bursts~\cite{zburst}. Any further constraint on UHE neutrino flux implies a
constraint on some parameter space of DM. Conversely, a narrow spectral feature
in the UHE flux could indicate the presence of dark matter, potentially solving
one of the major mysteries in modern physics. Especially at the highest
energies, PUEO is the only instrument that can probe this phase space.

\subsection{Fundamental Particle Physics with PUEO} 

UHE neutrinos detected by PUEO  would represent the
most energetic neutrinos available experimentally, allowing PUEO to probe fundamental
physics at a new scale. Observing the angular distribution through the earth
would allow a measurement of the neutrino-nucleon interaction cross-section~\cite{Connolly:2011vc,Klein:2013xoa}, which is sensitive to physics
beyond the Standard Model~\cite{Marfatia:2015hva,AlvarezMuniz:2001mk} and the
nucleus at small scales~\cite{CooperSarkar:2011pa}, in regions of parameter
space that are inaccessible by the Large Hadron Collider.  We expect that once
events are observed, PUEO could loosely constrain cross-sections at
$\sim 100$~TeV center-of-mass energies based on the energy-dependent zenith
angle distribution of the
events~\cite{Connolly:2011vc,Klein:2013xoa,Aartsen:2017kpd,Bustamante:2017xuy,Aartsen:2018vez}.
Like ANITA, PUEO will also be able to place constraints on Lorentz invariance
violation~\cite{anitaliv,Hooper:2005jp}. 

ANITA has detected several events in the air shower channel that, if 
not some unaccounted background, may require new physics explanations~\cite{anita1me,anita3me,
anita4cr}. PUEO will be able to search for more of these events with improved
sensitivity.  Specific search channels proposed for seeking physics beyond the
Standard Model in ANITA will be targets of searches with PUEO~\bsmcitations.

\subsection{Ancillary Studies} 

While the main focus of PUEO is the identification and characterization of UHE
neutrinos, the elevated radio platform allows for a number of ancillary science
results. PUEO measures UHECRs through their air showers from a geometry unlike
other experiments, allowing PUEO to probe radio emission models, including some
that may be relevant to the interpretation of the anomalous ANITA
events~\cite{krijnsteven}. 
PUEO, like ANITA~\cite{hical1, hical2}, will also be able to measure ice
reflectivity properties on Antarctica, which is important to the interpretation
of apparent upward-going air showers~\cite{shoemaker2020, reflections_on_reflections}.

\section{The Radio Detection Technique} 

Building a competitive UHE neutrino detector requires effective detector
volumes of many cubic kilometers. Even gigaton-scale optical Cherenkov
detectors that use natural media, such as IceCube~\cite{IceCube2016,
Aartsen:2018vtx} or Km3Net~\cite{km3net}, are of insufficient size to measure
the expected flux at the highest energies. Cost-effective scale-up is limited
by the $\mathcal{O}(10-100 \rm{m})$ effective length scale propagation of light
in their respective media. 

To reach the necessary detector sensitivities, it is advantageous to use a
technique profiting from the improved propagation properties of radio
frequencies and the very large visible volume provided by a high-altitude
platform. Two complementary methods are relevant to PUEO: in-ice Askaryan
emission and geomagnetic emission from air showers, which are briefly reviewed
here. 

\subsection{In-Ice Askaryan Emission}

\begin{figure}[b] 
  \centering
    \includegraphics[height=2.7in]{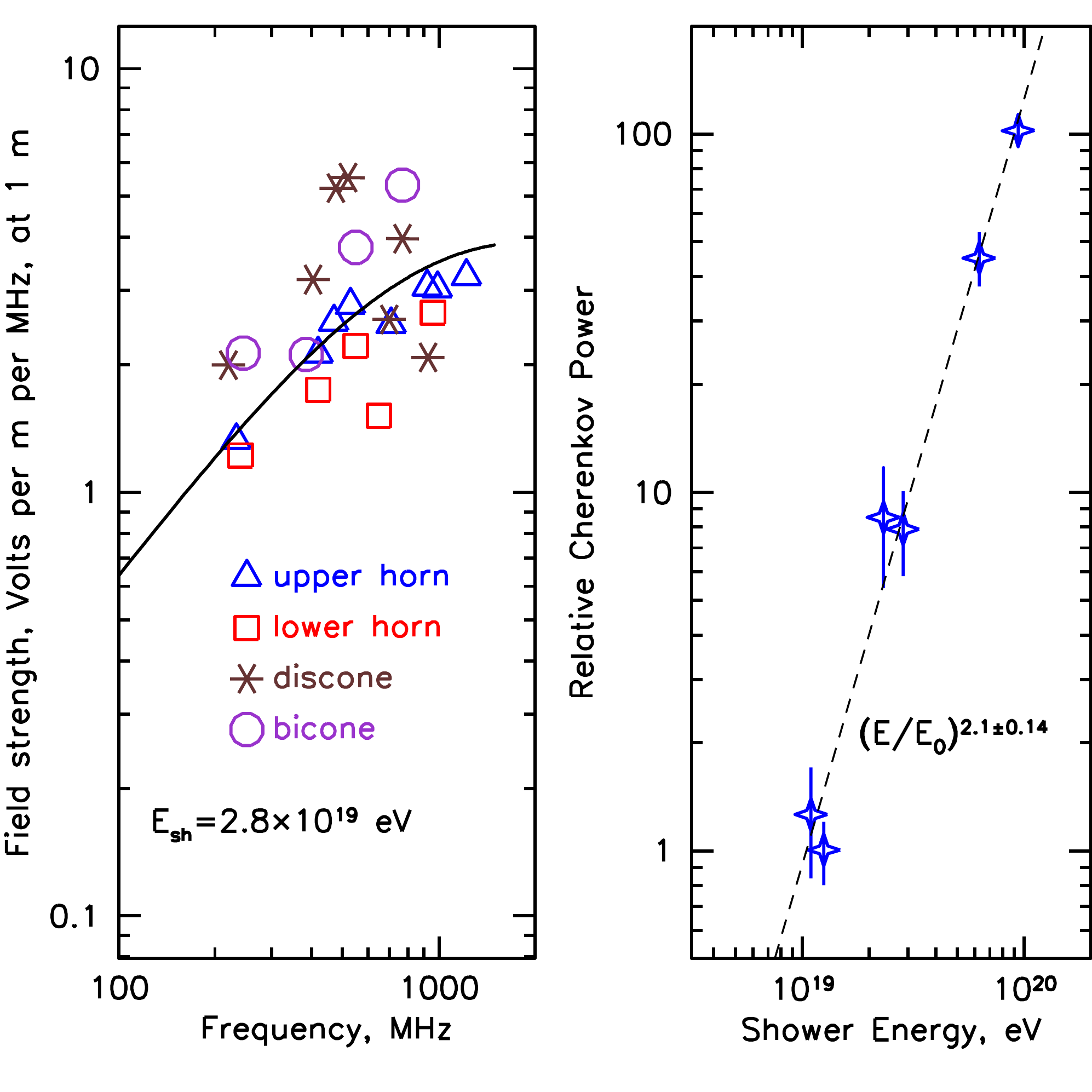}
\includegraphics[height=2.7in]{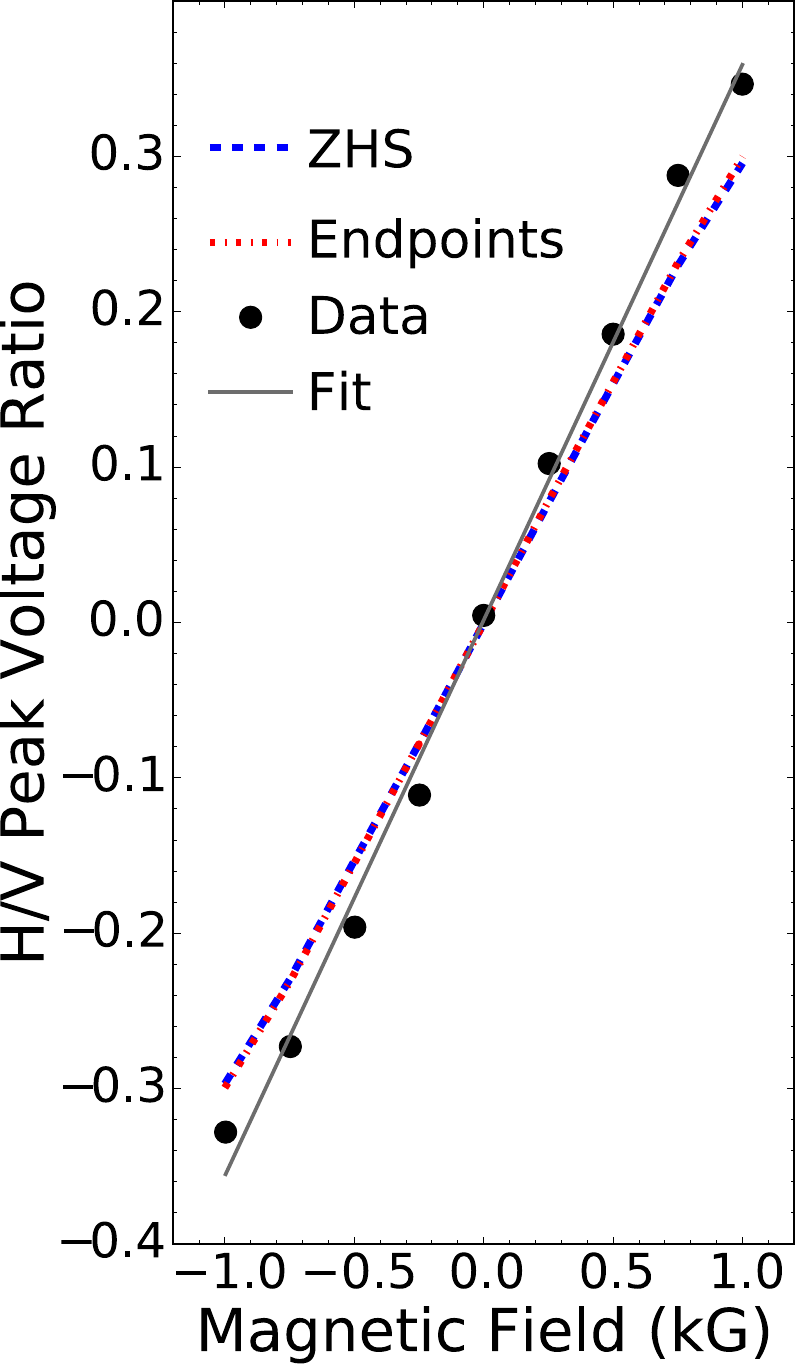}
    \caption{Key results from beam tests probing Askaryan and geomagnetic emission. \textbf{\textit{Left: }}Field strength vs. frequency of Askaryan 
  emission in the T-486 experiment. \textbf{\textit{Center: }}Coherence of the radiation detected. From Ref.~\cite{T486}. \textbf{\textit{Right:}} Horizontally-polarized signal fraction showing the 
expected linear behavior vs. magnetic field, from the T-510 experiment. From Ref.~\cite{t510}.}
    \label{fig:beamtest}
\end{figure}

The Askaryan Effect~\cite{Askaryan}, first predicted in the 1960's, refers to
the coherent radio emission resulting from an electromagnetic (EM) shower in a
dielectric medium.  Due to interactions with atomic electrons in the medium,
the EM shower develops a fast-moving negative charge excess on the order of
20\%. This manifests itself as coherent Cerenkov radiation at wavelength-scales
larger than projected size of the shower, from the point of view of the
observer. As the electric field scales with the shower energy, at high
energies, the radio emission becomes significant and is detectable by 
suitably-placed antennas, provided that the medium is sufficiently
radio-transparent. As hadronic showers quickly produce EM
showers, both hadronic and EM cascades 
produce emission via the Askaryan effect. UHE neutrinos interacting in dense media may induce either
type of shower (or both, when including secondary showers). 

A series of beam test experiments~\cite{Saltzberg:2000bk,Gorham:2004ny,Gorham:2006fy} have
demonstrated the Askaryan effect in various natural media, including salt,
silica, and ice. Of these, ice is the most abundant in the form of the glacial
ice in the polar ice sheets, and sufficiently-cold ice has km-scale radio
attenuation lengths~\cite{Barwick05}, making it a viable detection medium for
the Askaryan channel. 

T-486, which directly probed the Askaryan effect in ice, was performed in the SLAC End Station A (ESA) facility in 2006. 
A target of ice was constructed from close-packing rectangular blocks to form a 7.5 metric ton stack with dimensions $2 \times 1.5 \times 5$~m. 
The upper surface of the ice was carved to a slope of $\sim 8^{\circ}$  
in the forward direction %
to avoid total-internal reflection (TIR) of the emerging Cherenkov radiation at the
surface. 
Particle showers in the ice were produced by 28.5~GeV electrons in 10~ps 
bunches of typically $10^9$ particles,
with about 90\% of the shower contained in the 
target. %
The typical composite energy of the resulting showers is 
$3 \times 10^{19}$~eV, centered in
the cosmogenic neutrino range. 
Such composite showers can be used to validate the behavior of
radio emission from high-energy showers, with proper scaling. 
The ANITA payload, then an array of 32 dual-polarization, quad-ridged horn antennas sensitive between 200-1200~MHz, was used to 
receive the emission $\sim$15~m away from
the center of the target. %

Fig.~\ref{fig:beamtest} (left) shows measurements of the absolute field strength in several antennas. %
The uncertainty in these data are dominated by systematic errors 
and are $\pm40$\% in field strength.
The field strengths are compared to a parametrization based
on shower+electrodynamics simulations for ice~\cite{ZHS92, Alv97}, and the agreement
is within experimental errors. 
Fig.~\ref{fig:beamtest} (center)  also shows that the scaling of the pulse power with
shower energy is consistent with the expected quadratic behavior for coherent radiation.
Additional measurements made in the T-486 experiment were consistent with the geometry and expected polarization of the Cherenkov cone.

These data provide both a validation of the Askaryan effect
in ice, as well as a calibration of the response of detectors 
to an actual Askaryan signal. 
Since potential measurements of cosmogenic neutrino fluxes may not be
easily validated or confirmed by any other current experiment, at least in the short term, the SLAC T-486 beam test represents a major milestone in establishing
credibility for any possible discovery by a radio experiment.

In-ice Askaryan emission has also been studied in detail via simulations of
showers in ice~\cite{ARZ,ARZ2020}.   The Askaryan emission in ice is polarized
radially from the shower axis, with a broad emission cone that is centered at
the in-ice Cerenkov angle. Measurement of the polarization vector, even at just
one point, can therefore be used to constrain the shower axis and therefore the neutrino direction to $\mathcal{O}(15 (^\circ)^2)$~\cite{anita3source,anita3}. 

\subsection{Geomagnetic Emission from Air Showers}

Extensive air showers (EAS), normally the result of UHE cosmic rays (UHECRs)
interacting in the upper atmosphere, also produce radio emission. While there
is an Askayran component to these showers, the primary emission mechanism is
due to charge separation by the Earth's magnetic field~\cite{geomagnetic}. This
phenomenon is now well understood and radio detection of UHECRs is 
considered a  mature field~\cite{Hoover:2010qt,Schellart:2014oaa,Aab:2016eeq, Aab:2014esa}. 

The nature of geomagnetic emission has been probed both by detailed simulation efforts~\cite{Huege:2013vt, AlvarezMuniz:2011bs}, as well as beam test experiment T-510~\cite{t510}. The T-510 experiment took place in ESA at SLAC in 2014. Bunches of electrons with energy 4.35~or 4.55~GeV passed through 2.3 radiation lengths of lead pre-shower and entered a 4~m long plastic target, generating compact showers with a total energy equivalent to 
a $\sim4\times10^{18}$~eV cosmic-ray air shower.
The particle showers developed in high-density polyethylene, so the magnetic field and the frequencies of interest are scaled compared to those for EASs. 
Four %
ANITA horn antennas~\cite{ANITA-inst} 
recorded the electric fields radiated from the particle shower.

Fig.~\ref{fig:beamtest} (right) shows that the amplitude of the horizontally-polarized emission is 
linearly dependent on the magnetic field, as expected, and within the 20\% systematic uncertainty of 
the same slope as the predictions from simulation. The polarity of this induced voltage changes 
sign when the direction of the magnetic field flips direction, indicating that the 
transverse current flows in the opposite direction. The vertically-polarized emission 
is observed to be constant with respect to magnetic field strength.  
The angular radiation pattern as a function of frequency was also measured and is consistent with simulations. %

For PUEO and other neutrino detectors, the main interest in EAS is that they
provide another means through which to detect neutrinos using radio. A tau
neutrino, interacting in the Earth or ice via a charged-current interaction,
will produce a tau lepton that can escape into the atmosphere and subsequently
decay, producing an apparent upward-going EAS ~\cite{airshowerchannel}, which
can then be detected via its radio emission. Upward-going air showers may be
distinguished from reflected downward-going showers via the phase structure of
the electric field, as reflected air showers undergo phase inversion from the
air-ground interface~\cite{anita1me,anita3me}. 

The EAS $\tau$ channel typically has a lower energy threshold than the Askaryan
channel. Due to the narrow emission cone, the overall aperture is less once the
Askaryan channel threshold is exceeded. However, the small opening angle of the in-air cone
gives the EAS channel improved pointing resolution $\mathcal{O}(1^\circ)$  compared to the Askaryan channel. 

\subsection{Radio Detection Platforms} 

A number of existing experiments have employed radio detection techniques for
detection of UHE neutrinos. Experiments such as
ARA~\cite{ara_performance,ara_2station} and ARIANNA~\cite{arianna15b} have deployed
prototypes in Antarctica seeking to measure Askaryan emission from neutrinos
interacting in nearby ice. Future experiments such as RNO~\cite{RNO_Whitepaper}/RNO-G~\cite{RNO-G} and IceCube
Gen2 Radio ~\cite{Gen2Astro2020WhitePaper} plan to deploy even larger arrays.
BEACON~\cite{beacon}, TAROGE~\cite{taroge} and GRAND~\cite{grand} are proposed designs to detect
radio emission from tau-neutrino-induced air showers in mountainous regions. 

Compared to ground-based detectors, the balloon platform used by ANITA and
PUEO provides a much larger instantaneous volume, as more target volume
is directly visible from the payload. However, this larger
aperture generally comes at the price of a higher threshold, as the visible
events are farther away. ANITA has set the best limits on UHE
neutrinos at the highest energies ($\gtrsim$ 30 EeV)~\cite{anita4}. 
The greatly reduced trigger threshold available with PUEO allows balloon-borne searches for UHE neutrinos to be 
competitive with limits from ground-based experiments at energies below 30 EeV. Due to their large
instantaneous aperture, balloon-based experiments are also ideal for searching
for transients, which might produce multiple events at a time that could exceed
background levels only in detectors with larger instantaneous sensitivity.

\section{Review of Results from ANITA} 
\label{sec:anita} 

\begin{figure}
    \centering
    \includegraphics[width=4in]{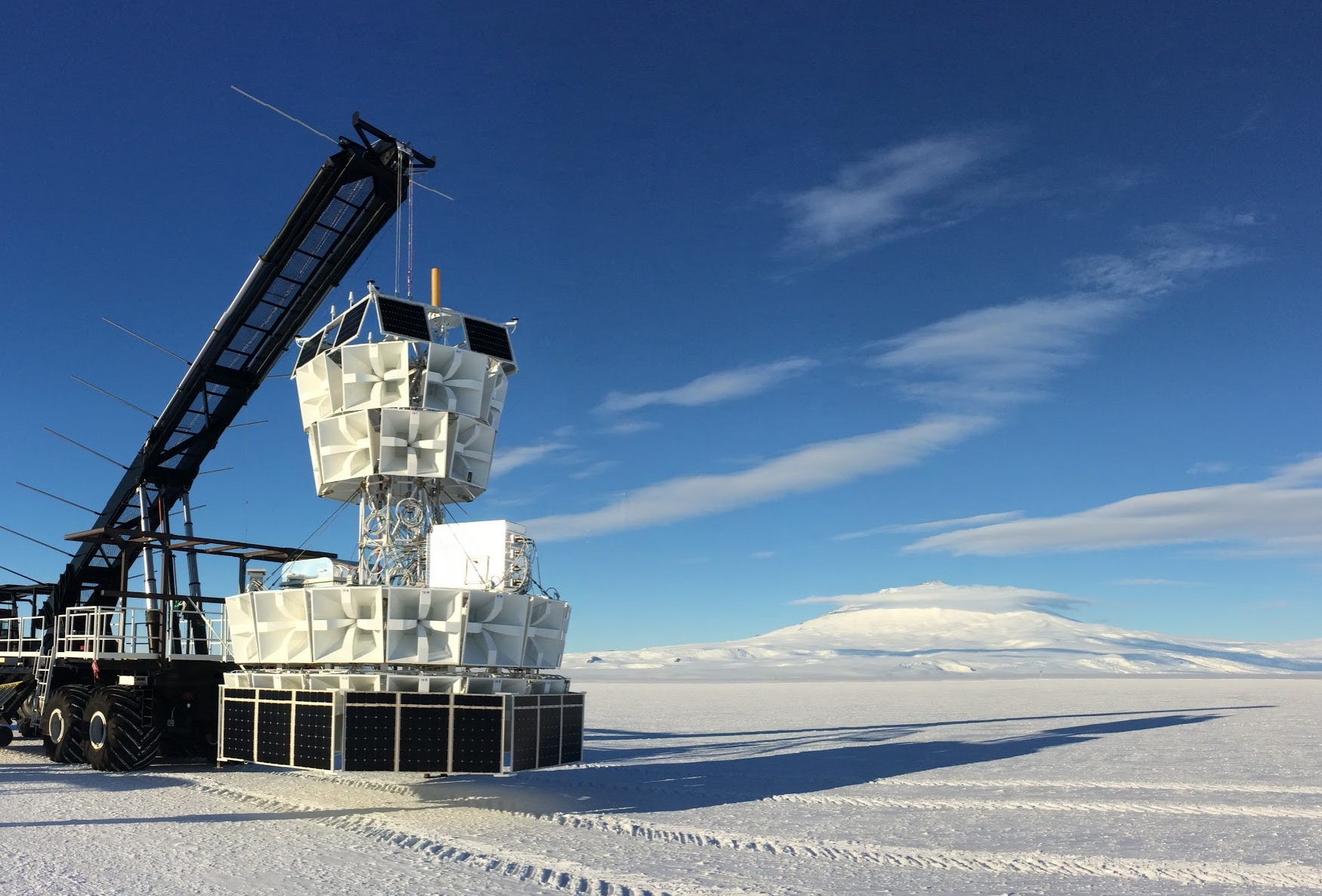}
    \caption{A picture of ANITA-IV, preparing for the 2016 launch on the Ross Ice Shelf in Antarctica.}
    \label{fig:anitaPicture}
\end{figure}

Before detailing the many improvements in PUEO, we briefly review results from the predecessor ANITA program, which completed four successful flights between 2006 and 2016. ANITA (Fig.~\ref{fig:anitaPicture}) is the pioneer in balloon-borne radio instruments, having set world-leading limits on diffuse UHE flux and detected dozens of UHECR air showers.

\subsection{ANITA Constraints on UHE Neutrino Flux}

\begin{figure}
\centering
\includegraphics[width=\textwidth]{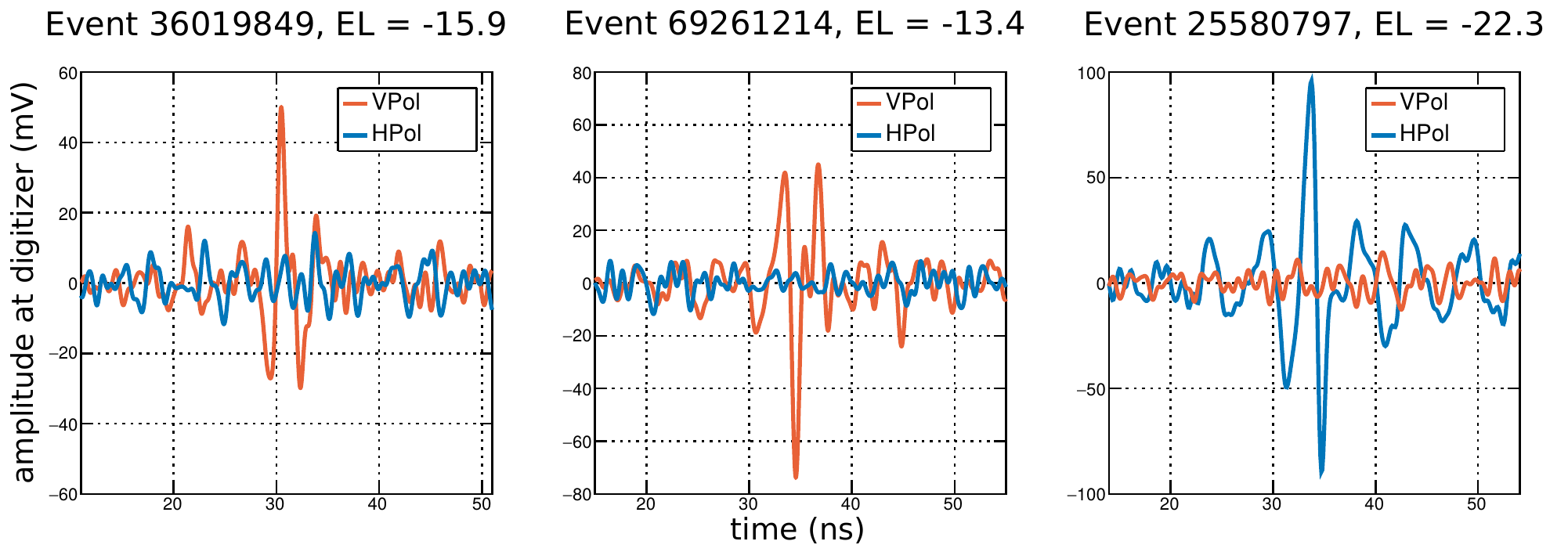}
\caption{\textbf{\textit{Left: }} The single vertically-polarized candidate from Analysis~\textbf{A} in ANITA-IV.  \textbf{\textit{Middle: }}The single vertically-polarized candidate from Analysis~\textbf{B} in ANITA-IV.  \textbf{\textit{Right: }}An example cosmic-ray candidate from ANITA-IV.} 
\label{fig:a4nuresults}
\end{figure}

ANITA was designed to look for Askaryan emission from UHE neutrino-induced showers in Antarctic ice, and has placed the best constraints on the UHE neutrino flux between $10^{19.5}$~eV -- $10^{21}$~eV.  Each ANITA flight has published constraints on this flux~\cite{anita1,anita2,anita3,anita4}, with modest improvements in sensitivity.  With each flight, analysis techniques have improved, and multiple independent analysis pipelines have been developed.  The most recent flight (ANITA-IV) saw two independent analyses, one with the highest analysis efficiency yet for ANITA: $82\pm2\%$, after constructing cuts to isolate Askaryan neutrino candidates.  The blind analysis techniques used for these analyses are sophisticated, robust, and have produced consistent results over the four flights of ANITA. Fig.~\ref{fig:a4nuresults} shows the results of the two independent blind searches for Askaryan signals in the ANITA-IV data.  Each analysis had one candidate event survive on background estimates of $0.64^{+0.69}_{-0.45}$ and $0.34^{+0.66}_{-0.16}$, respectively.  The dominant contribution to background was difficult-to-reduce man-made backgrounds.  The combined published limit from ANITA I-IV is shown in Fig.~\ref{fig:current_limits}.

In addition to diffuse searches for UHE neutrino flux, ANITA has also specifically set limits on fluences from astrophysical sources such as GRBs, flaring blazars, supernovae, and IceCube-identified sources ~\cite{anitaGRB, anita3source}. The ANITA-III source search establishes a general methodology for searching for UHE neutrinos in spatial and temporal coincidence with source classes, and furthermore identified the surviving diffuse ANITA-III event as potentially associated with SN 2014D, although not at a statistically significant level. 

Over time, ANITA has seen more radio-frequency interference (RFI) from satellites, as more satellites are launched.  The change in RFI environment between ANITA-II and~-III was especially apparent in the data, and caused ANITA-III to lose livetime and therefore sensitivity.  With each flight, ANITA has made changes to adapt to the changing RFI environment. ANITA-IV implemented tunable notch filters~\cite{TUFF} that allowed us to dramatically improve the fractional livetime of the instrument (from 32\% in ANITA-III to 91\% in ANITA-IV). However, for nearly all of the flight, most of the power below 300~MHz needed to be filtered out, in order for the instrument to continue to take data with low trigger thresholds, which also cost some sensitivity to neutrinos and cosmic rays.

\subsection{ANITA Cosmic Ray Measurements}
ANITA-I made the serendipitous first detection of UHE ($>10^{18.5}$ eV) cosmic rays via geomagnetic radio emission from the EASs that cosmic rays create in the atmosphere~\cite{Hoover:2010qt, schoorlemmer}. Since then, significant progress has been made in simulations, analysis, and with the T-510 beam test, in understanding this radio emission mechanism.

Geomagnetic emission from EASs in the ANITA data is distinct from Askaryan emission from neutrinos interacting in the ice due to differences in observed polarization.  In Antarctica, where Earth's magnetic field is primarily vertical, radio signatures from EASs will appear predominantly horizontally polarized.  Askaryan signals from neutrinos interacting in the ice will produce predominantly vertically-polarized events, due to Earth absorption of up-going UHE neutrinos, the polarization of the Cherenkov cone, and preferential transmission at the ice-air interface.

There are two ways for ANITA to see UHE cosmic rays.  Since ANITA sits above most of the atmosphere and cosmic rays are down-going as viewed from the Earth, most of the cosmic-ray events that ANITA sees are from radio signals that are reflected off of the Antarctic surface, and then come back up to ANITA.  
A small fraction of cosmic-ray events observed by ANITA are from cosmic rays that, as viewed by ANITA, are slightly upgoing, skimming through the atmosphere at elevation angles above the horizon (at $\sim-6^\circ$ elevation viewed by ANITA), but below the horizontal (since there is negligible atmosphere above ANITA).  For events in which the radio signal reflected off of the surface, an polarity inversity of the signal is expected compared to radio emission that is observed directly from the cosmic-ray air showers that come from above the horizon, resulting in the two categories of ``direct" and ``reflected" events.

\begin{figure}
\centering
\includegraphics[height=2.5in]{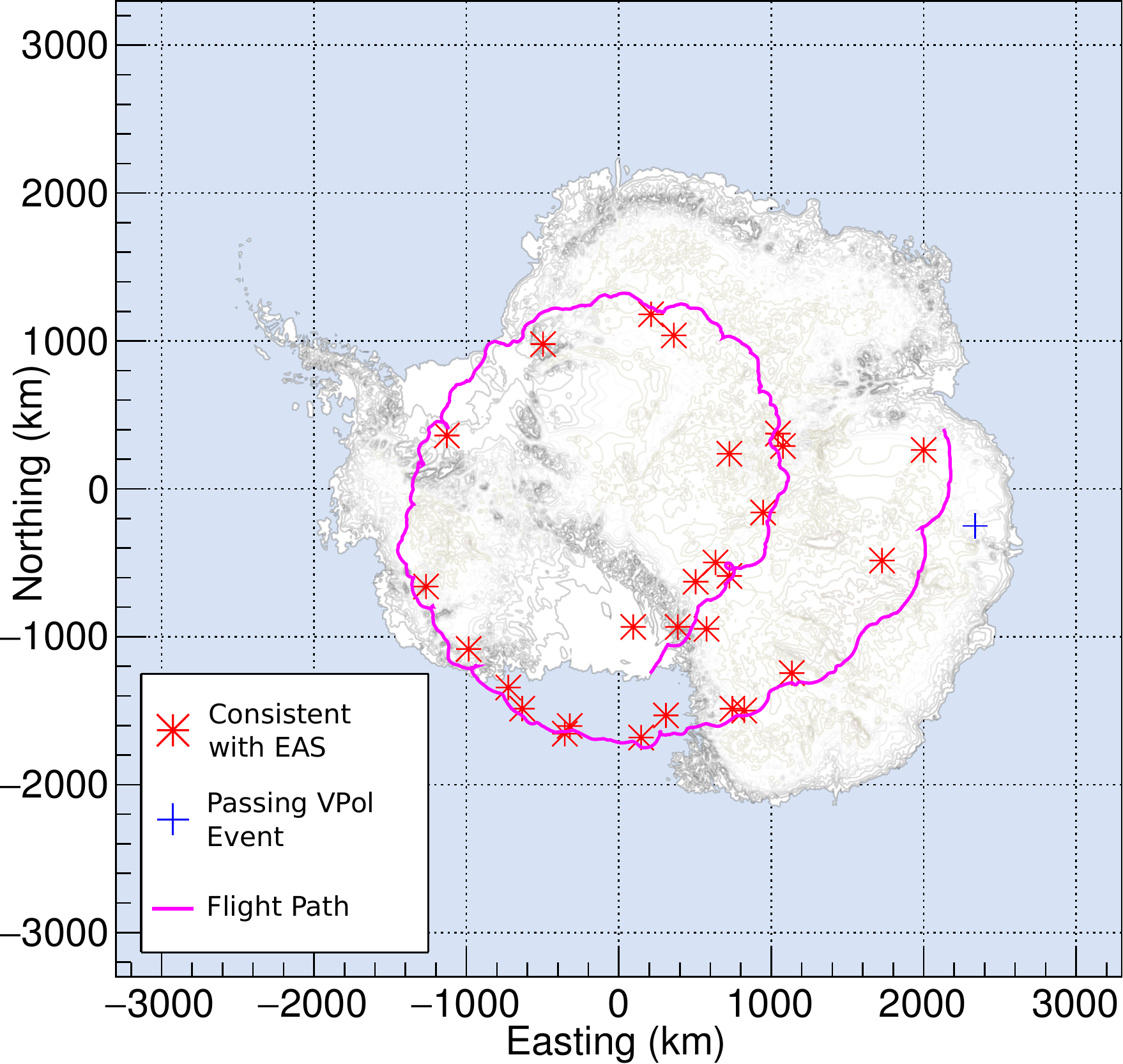}
\includegraphics[height=2.5in]{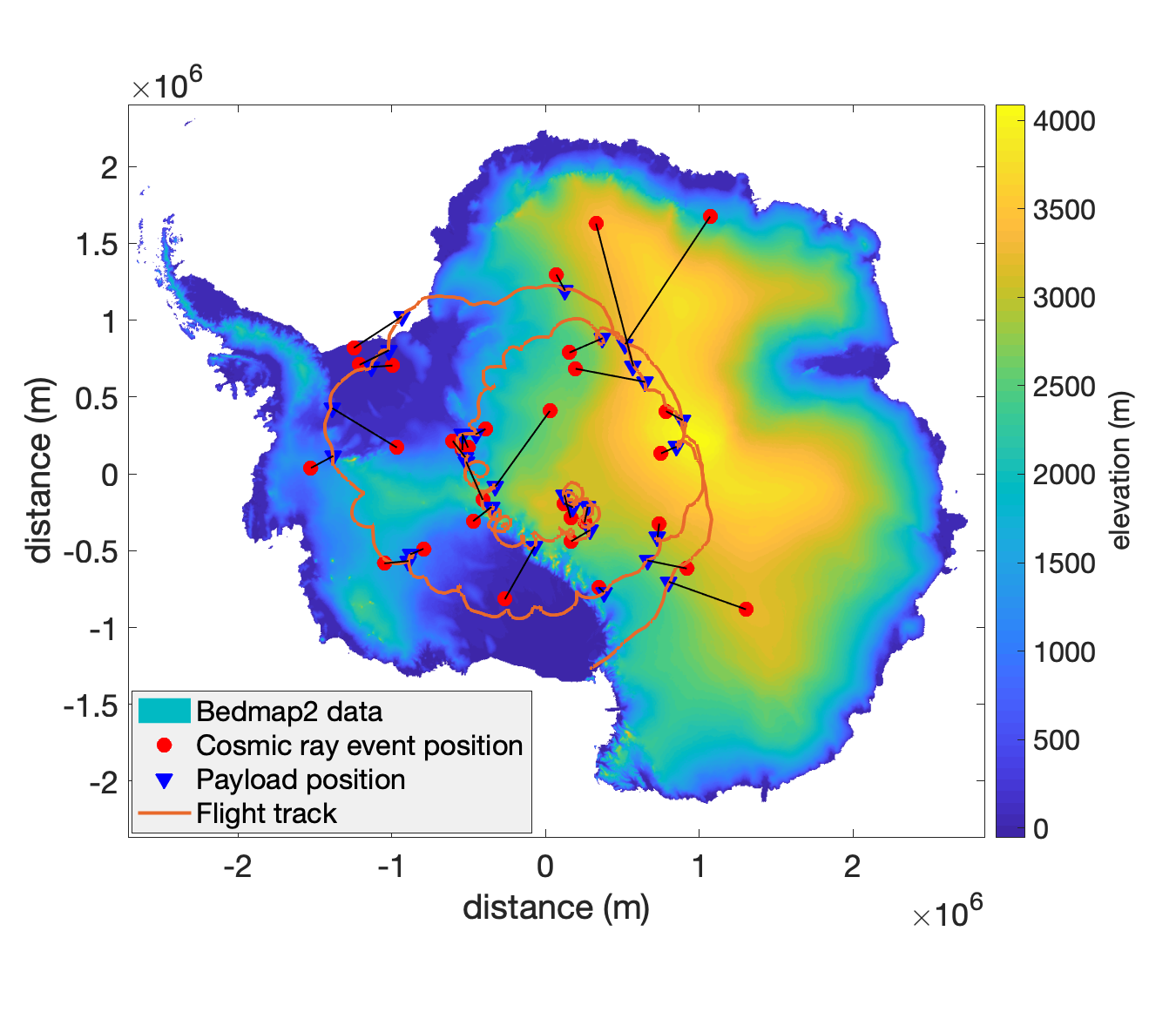}
  \caption{\textbf{\textit{Left:}} A map of EAS events and the 
  candidate Askaryan neutrino event from the most sensitive analysis of ANITA-III. From Ref.~\cite{anita3}. \textbf{\textit{Right:}} A map of EAS candidates from ANITA-IV. From Ref.~\cite{anita4cr}. }
\label{fig:a3crMap}
\end{figure}

After ANITA-I detected 16 cosmic-ray events, two of which were direct events and 14 of which were reflected events with opposite polarity, searches for EAS-like events became an additional science driver for ANITA.  Over the four flights of ANITA, there have been 71 %
events identified that are consistent with signatures of cosmic rays, on a combined background of order 1.  Seven of these events are direct signatures, and the other 64 are reflected events with opposite polarity.  Fig.~\ref{fig:a3crMap} shows a map of the identified EAS candidates from ANITA-III.  Figs.~\ref{fig:a4nuresults} and~\ref{fig:a1ME} show examples of cosmic rays detected with ANITA-I, -III, and -IV.

\subsection{ANITA's Upward-Going Cosmic-Ray-Like Events}
\label{sec:me}
ANITA has also observed a few anomalous cosmic-ray-like events.  ANITA-I,~-III and~-IV have seen an excess of EAS candidate events that come from below the horizon but do not have inverted polarity.  These have the same polarity as the observed direct cosmic-ray air showers but come from below the horizon.  ANITA-II was significantly less sensitive to the EAS channel, since its trigger was optimized solely for the Askaryan neutrino search (and at the time unknowingly optimized away from an EAS search). 

\begin{figure}[t]
\centering
\includegraphics[width=\textwidth]{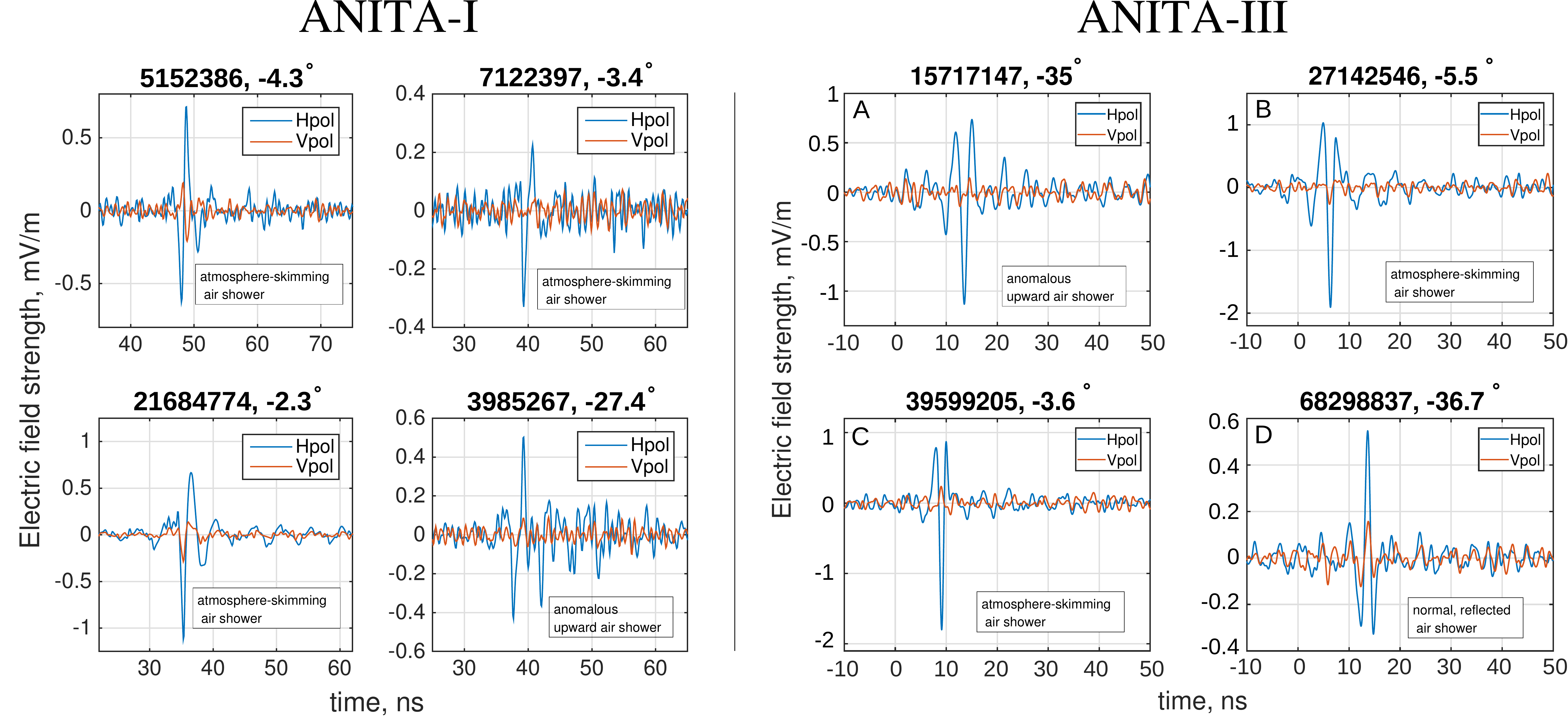}
\caption{\textbf{\textit{Left: }} ANITA-I: Waveforms for four of the EAS-candidate events, including one anomalous-polarity event in the lower-right panel. The waveform titles include the event number and the elevation angle relative to horizontal. The horizon is typicaly at -6$^\circ$, depending on altitude.  From Ref.~\cite{anita1me}. \textbf{\textit{Right: }}ANITA-III: Event waveforms four of the EAS candidate events, including one anomalous-polarity event in the upper-left panel.  From Ref.~\cite{anita3me}.} 
\label{fig:a1ME}
\end{figure}
ANITA-I and~-III each saw one anomalous-polarity event.  With the low backgrounds in this channel, these are $3.3\sigma$ and $3.0\sigma$ effects, respectively. %
Fig.~\ref{fig:a1ME} shows these events with anomalous polarity, compared to observed EAS events with those same payloads.  The left half of Fig.~\ref{fig:a1ME} shows four events from ANITA-I: three observed above-horizon direct cosmic-ray events, and the one anomalous-polarity event in the lower-right panel.  The right half of Fig.~\ref{fig:a1ME} shows four events from ANITA-III: the mystery event in the upper-left panel, the two above-horizon direct cosmic-ray events, and an example of a reflected cosmic-ray event with inverted polarity in the lower-right panel.

One physics explanation for non-inverted upward-going cosmic-ray-like events could be that they were created by tau neutrinos interacting in the Earth, creating a tau lepton, which then decays in the atmosphere and induces an EAS. However, both the ANITA-I and~-III anomalous-polarity events come from very steep elevation angles: $-27.4^\circ$ and $-36.7^\circ$.  This is inconsistent %
with limits from ground-based observatories on an isotropic tau neutrino flux under the assumption of standard-model cross sections~\cite{anitaIsotropicTaus}.  Recently, a novel interpretation of the signals
as transition radiation from the vanishing
of transverse currents in cosmic-ray showers at the air-ice
boundary has been proposed
~\cite{krijnsteven}. These events could be a hint of beyond the Standard Model physics: supersymmetric interpretations, sterile neutrino interpretations, and dark matter interpretations have all been put forward as potential hypotheses~\bsmcitations. 
Following up on these apparent upward-going cosmic-ray-like events from ANITA-I and~-III is a science priority that motivates the design of PUEO. The direction of emission from the ANITA-III anomalous event was consistent in position with recently-exploded SN 2014dz, raising the possibility of some sort of supernova-induced transient. 

\begin{figure}[t]
\centering
\includegraphics[width=0.67\textwidth]{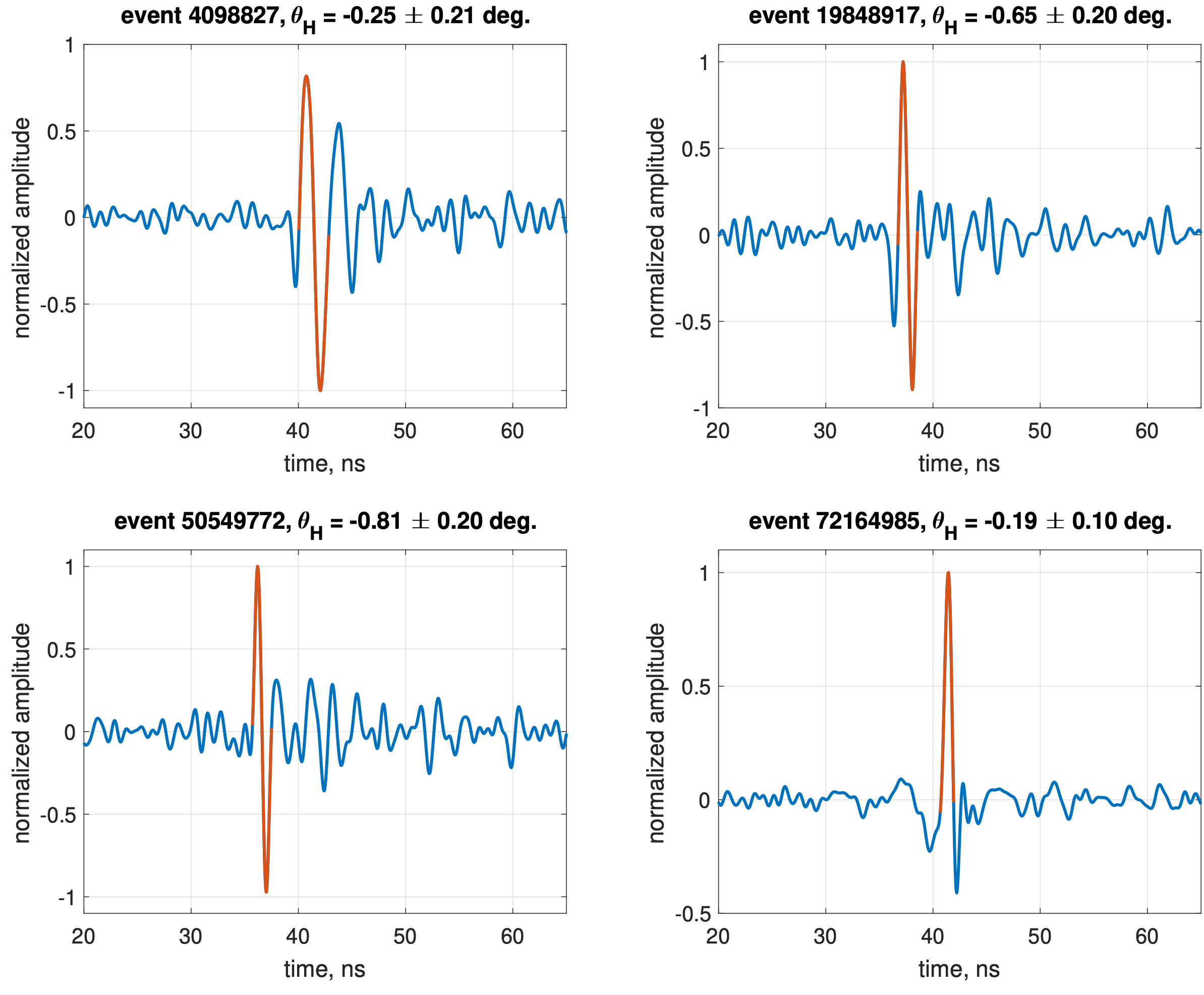}
\caption{The candidate upward shower events from ANITA-IV. Unlike previous flights, all upward showers are very close the horizon. Due to the varying instrument response during the ANITA-IV flight deconvolution is required to determine relative polarity. Here, the angle below the horizon is reported with each event, rather than the angle below horizontal. The horizon is typically around 6 degrees below the horizontal, but is different for each event due to variations in payload and ground altitude.} 
\label{fig:a4ME}
\end{figure}

Recently, ANITA has reported on the results of the upward-shower search in ANITA-IV
~\cite{anita4cr}. ANITA-IV identified four upward-shower candidates
(Fig.~\ref{fig:a4ME}). Unlike previous such events, these were all near the
horizon and each individual one has a non-negligible chance of being an
anthropogenic background or having been misidentified as from above the
horizon. As such, the total significance for at least one of the events not being some sort of background (anthropogenic, misidentified polarity, or misreconstructed on the wrong side of the horizon) is only 3.2$\sigma$. The ANITA-IV
search does not exclude the steep events from previous flights, but, if these new
events are truly from upward-air showers, they may be a new class of events.
Without such a large chord through the earth, there is no tension with the
standard model with the angular distribution of this set of events. The compatibility with
measurements from other experiments is still under study. PUEO will have
greatly improved sensitivity to this class of events, and, with better pointing
ability, additionally have a lower background.

\section{The PUEO Instrument} 

The PUEO payload will consist of a 216~channel Main Instrument (300-1200~MHz) and an 8-channel Low Frequency (LF) instrument, which will cover 50-300~MHz.
The overall concept of the PUEO payload is similar to that of ANITA. Much of the mechanical and RF design, power systems, attitude and location systems, and data storage and transfer is inherited from ANITA.  However, PUEO represents a significant improvement in sensitivity compared to the ANITA payload.  This is achieved by:

\begin{enumerate} 
  \item An interferometric phased array trigger, which lowers the trigger threshold, and increases the expected neutrino and cosmic-ray event rate. 
  \item More than doubling the antenna collecting area above 300~MHz.  This is enabled by increasing the low-frequency cutoff of the antennas from 180~MHz for ANITA-IV to 300~MHz for PUEO, which reduces the size of the antennas by a factor of two in area. 
  \item Adding a drop-down dedicated low-frequency instrument, as well as a downward-canted 12-antenna high-frequency dropdown. These dedicated instruments will greatly improve PUEO's sensitivity to air showers, over a range of elevation angles. 
  \item Significantly improved ability to filter man-made noise in real-time at the trigger level. 
  \item Significantly improved pointing resolution, especially in elevation, from a combination of better orientation measurements and a larger physical vertical baseline.  Improved elevation pointing resolution will allow us to improve analysis efficiency and reduce contamination from man-made backgrounds.
\end{enumerate}

Trade studies over the past several years have 
 explored a wide range of possible payload configurations, including a
 variety of antenna designs (frequency ranges, gain, physical size), antenna
 geometries (large collapsible drop-downs, azimuthally asymmetric
 configurations, various antenna spacings and orientations). PUEO is the
 optimal design that is currently achievable with technologies that have
  been developed and demonstrated that maximizes the achievable sensitivity
  given the power, mass, and size constraints of an LDB payload.

\begin{figure}[p]
 \centering
\includegraphics[width=3.5 in]{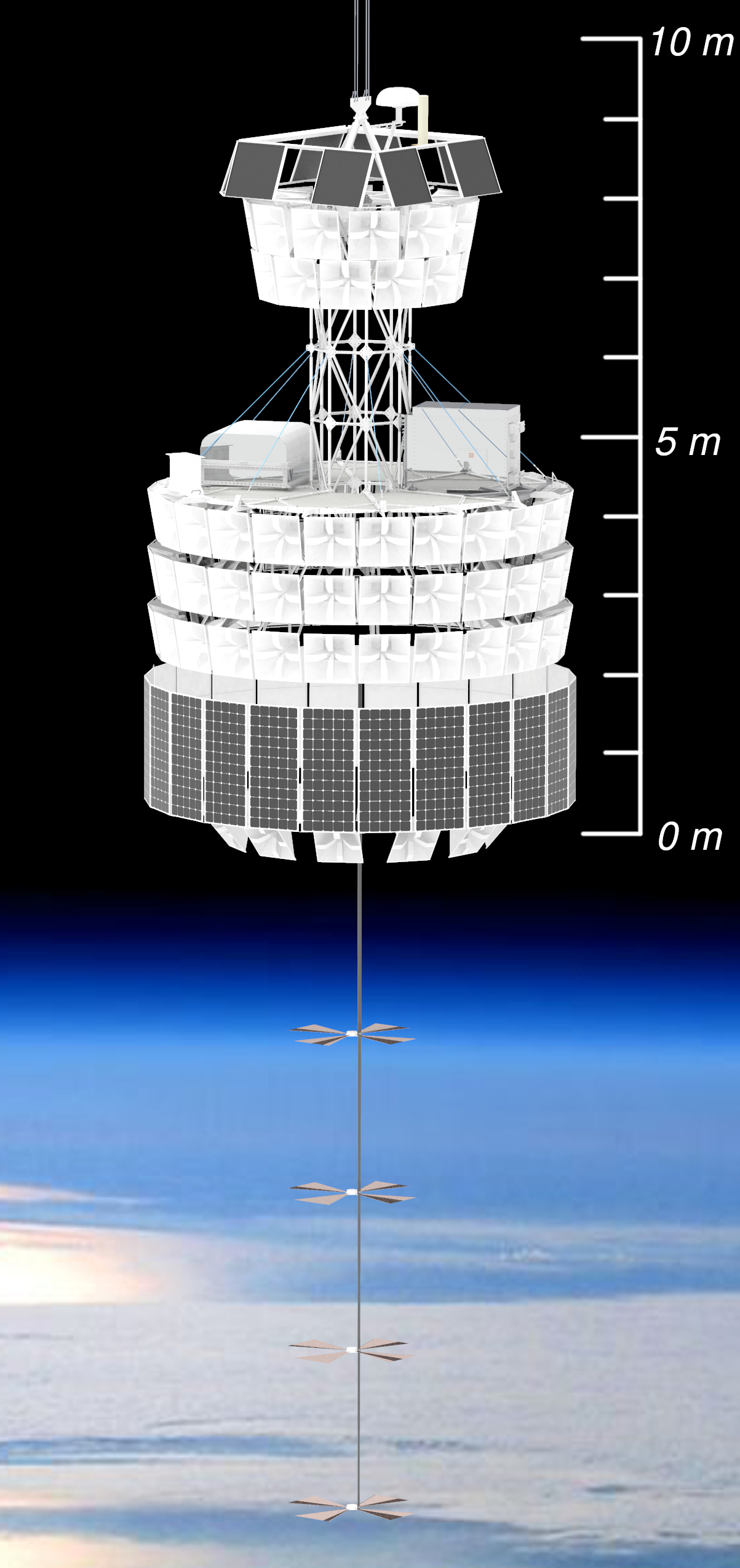}
\caption{A rendering of the PUEO payload, including a potential design for the low-frequency drop down.} 
\label{fig:rendering} 

\end{figure}

Fig.~\ref{fig:rendering} shows a rendering of the PUEO payload.  PUEO receives
radio signals from cosmic particles using its 108 dual-polarized
quad-ridged horn antennas, sensitive between 300~MHz and 1200~MHz in the Main Instrument, and 8 LF antennas. Radio
signals observed by these antennas are amplified, digitized above the Nyquist
frequency, and a trigger decision is made in real time to determine which data
are saved to disk.

\subsection{Maximizing Sensitivity Using an Interferometric Phased Array Trigger} 

An interferometric phased array trigger will be employed for the Main Instrument of PUEO, a technology that has been pioneered through the work of multiple groups in PUEO, and has successfully been demonstrated {\it in situ} at the South Pole on the Askaryan Radio Array (ARA) experiment~\cite{oberla}, achieving the lowest demonstrated trigger threshold in any radio detector for cosmic neutrinos.  
The key to lowering the trigger threshold (and consequently lowering the energy threshold and increasing sensitivity at higher energies) is the ability to distinguish weak neutrino-induced impulsive signals from thermal noise.  
The phased array trigger coherently sums the full radio waveforms with time delays corresponding to a range of angles of incident plane waves, averaging down the uncorrelated thermal noise from each antenna while maintaining the same signal strength for real plane-wave signals (such as neutrinos). 
This gives a boost in 
signal-to-noise ratio (SNR) for triggering that goes as $\sqrt{N}$, where $N$ is the number of antennas included the trigger array.  
Such interferometric techniques have been extensively used in radio 
astronomy~(for a review, see~\cite{thompson}).

The interferometric trigger also provides improved rejection of man-made RF interference,
which tends to come from localized directions.  %
At any given time, the beams that correspond to directions where there is man-made interference may be masked, further improving detector performance.

\subsection{Antennas and Radio Frequency Signal Chain for the Main Instrument}
\begin{figure}[p]
    \centering
    \includegraphics[width=\textwidth]{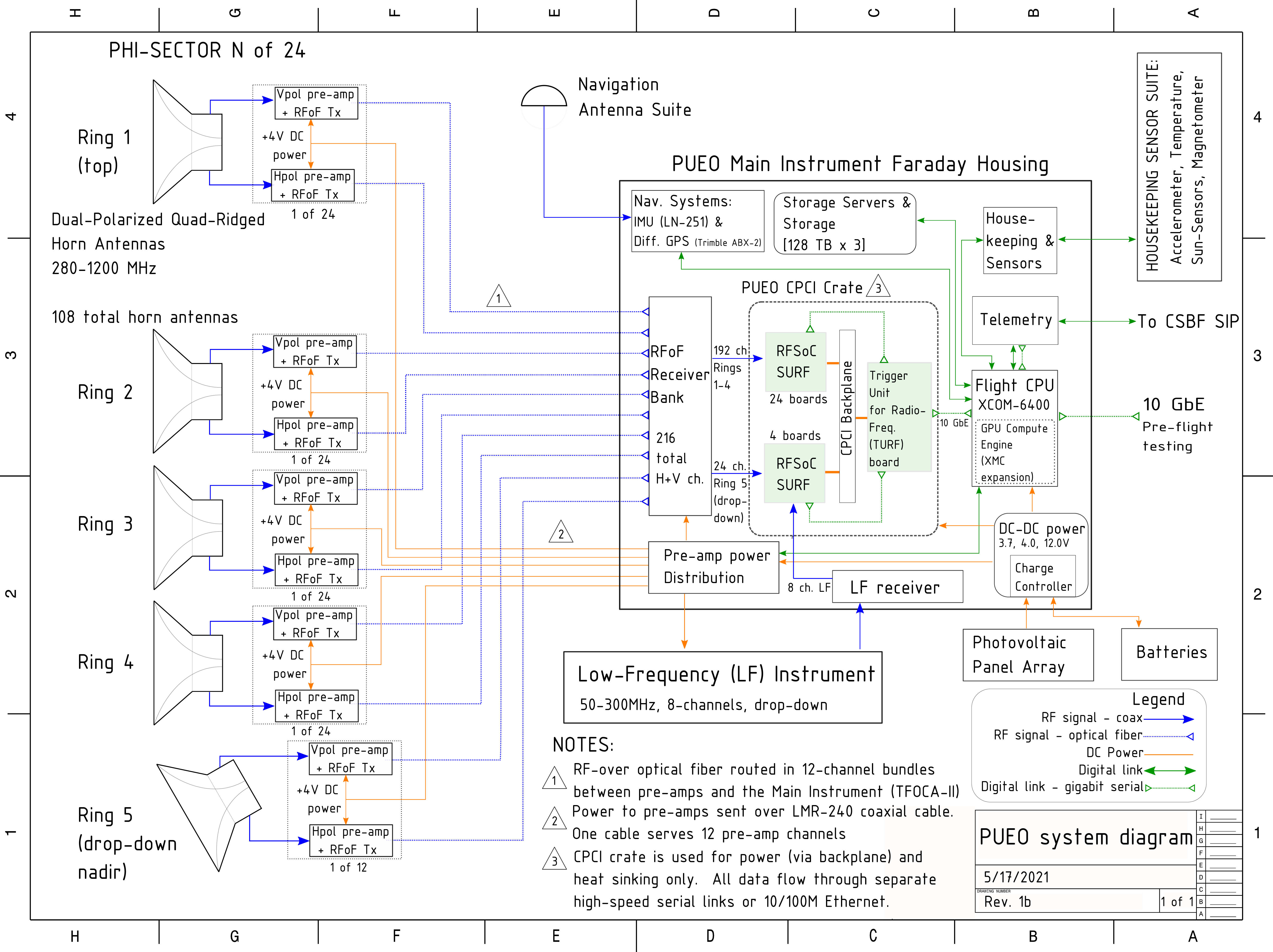}
    \caption{The PUEO system diagram.}
    \label{fig:pueo_diagram}
\end{figure}

PUEO's Main Instrument will use dual-polarization quad-ridge horn antennas designed and manufactured by
Antenna Research Associates. These antennas have a bandwidth of
300-1200~MHz with a physical aperture that is half of the ANITA-III and~-IV horns, which
had a lower frequency limit of $\sim$200~MHz.
PUEO will be instrumented with 108~horn antennas in 5~payload rings with 24~azimuthal `phi-sectors' as shown 
in Figs.~\ref{fig:rendering} and~\ref{fig:pueo_diagram}, a significant increase over the 
48-antenna ANITA-IV payload and more than doubling the effective receiving area at 300~MHz. Each `phi-sector' consists of the set of 4-5 antennas aligned to the same azimuth.  Antennas on the upper four rings are canted at $10^\circ$ below the horizon. The lowest 12~horn antennas will drop down after launch, and will be canted downward at a $40^\circ$ angle, providing additional sensitivity to EAS events.   The antennas each have a field of view of $\sim\pm30^\circ$, so each signal is viewed with multiple azimuthal sectors of antennas on the payload.
In part, the 300~MHz cutoff
is chosen to reject satellite continuous wave (CW) and other RFI
that made large portions of the 200-300~MHz band unusable during the the ANITA-III and~-IV flights. The relative sensitivity hit from losing the remaining non-RFI-contaminated bands below 300~MHz is relatively small ($\sim20-30$\%) and is more than compensated for by the lower threshold afforded by use of additional antennas.

The received signal at each antenna is amplified by $\sim60\,\text{dB}$
using a cascaded amplifier chain that includes low-noise amplifiers (LNAs) mounted directly at the antenna
in RF-tight pre-amplifier enclosures as shown in Fig.~\ref{fig:pueo_diagram}. 
We expect to achieve an overall system temperature, $T_{sys}$, of $\sim$160~K, which is dominated by the temperature and fraction of ice in the field of view of PUEO (a contribution of $\sim$110~K). The PUEO electronics contribute an additional $\sim$50~K across the band using a mature design with ANITA heritage,
which gives an overall $\sim5\%$ improvement to the total system temperature compared to ANITA-IV.  
The frequency bandpass is defined
using high-order filters to reject out-of-band RF interference. 
We choose to use commercially-available RF-over-fiber (RFoF) transceivers and bundled 12-channel tactical cables 
for signal transport to a central instrument box, rather than using single-channel 
coaxial cable, for the significant reduction in cabling complexity for the 224~channels in PUEO.
RFoF also critically reduces the cable mass of the payload while more than doubling the channel count over the ANITA-IV mission. High-frequency transmission losses between the front-end LNA and the instrument are also minimized by using the RFoF. These commercial RFoF links have been previously qualified in thermal-vac tests by ANITA. 

\begin{figure}[p]
    \centering
    \includegraphics[width=\textwidth]{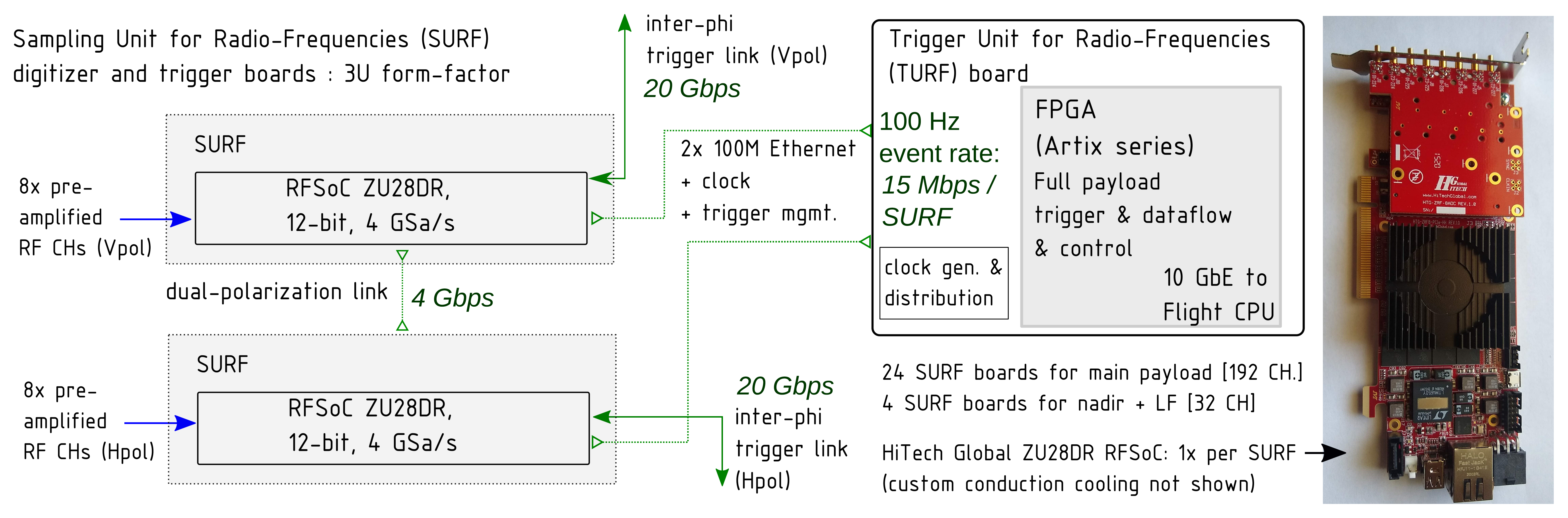}
    \caption{Diagram of the PUEO data acquisition and trigger system.}
    \label{fig:pueo_daq}
\end{figure}

\subsection{Triggering and Data Acquisition (DAQ) System}

The full 224~RF channels in the combined main instrument an LF~instrument  are
converted to RF by a bank of RFoF receivers 
and connected to the digitizing $\&$ triggering system crate, which consists 
of fourteen 16-channel Sampling Unit for RF (SURF) boards
and a master trigger and data collection unit termed the Trigger Unit for RF (TURF)
as shown in Fig.~\ref{fig:pueo_daq}. The TURF and SURFs are connected via standard 100Mbps Ethernet links and a set of point-to-point low-latency control signals.
The TURF also provides a global clock to each SURF board, ensuring all RFSoC digitizers
are synchronized.
The PUEO DAQ system is housed in a custom 3U crate ($\sim$18cm$\times$18cm$\times$70cm)
used for power distribution and conductive heat dissipation. 

{\bf Xilinx Radio-Frequency System-on Chip (RFSoC):}
The Xilinx RFSoC is a relatively new technology that incorporates
fast, high-resolution analog-to-digital converters (ADCs) 
and digital-to-analog converters (DACs) within
the same chip fabric as a high-performance field-programmable array (FPGA)~\cite{xilinx_rfsoc}. This drastically reduces the 
digitizer power consumption by eliminating the need
for driving fast digital data lines off-chip. 
This lower power, combined with the newly-available commercial RFSoC boards with extended-temperature grade (rated to 100$^{\circ}$C), makes this technology well suited as a combined readout and interferometric trigger system for PUEO. 

{\bf RFSoC SURF: } The SURF board holds one commercially-available HiTech Global RFSoC board running at 4~GSPS with 12 bit resolution, as shown in Fig.~\ref{fig:pueo_daq}. The SURF connects as a 
conduction-cooled board stack and fit in a single 3U slot in the custom DAQ crate.
Pairs of neighboring SURFs are connected with a 4-Gbps link to merge the polarization data 
from a set of 8 dual-polarization horn antennas, allowing real-time polarization measurements as a trigger discriminant for linearly-polarized Askaryan and air-shower signals.
Each of these `SURF pairs' manages the filtering, data buffering, and triggering for a 2-phi sector azimuth region of the main payload. The drop-down horns and LF instrument together require four additional SURF boards (two SURF assemblies), which will be programmed with a separate trigger. The high-speed DACs and DDR4 memory interface on the RFSoC are not used and are powered off, resulting in an estimated power consumption of $\sim$17~W per 8-channel RFSoC module.

The digital data on the RFSoC FPGA are split between a `trigger'
and `recording' path. The data sent to the recording buffer are saved as-is for analysis, and can be stored in real-time up to a length of 10~microseconds to accommodate the trigger latency.
The trigger path includes a set of filtering, beamforming, and real-time signal de-dispersion to maximize the PUEO trigger sensitivity.  The first stage of the PUEO trigger will include
a dynamic digital notch filtering of two known
RFI bands (375 and 460~MHz), supplanting the analog notch filters used on ANITA-IV~\cite{TUFF} to successfully mitigate satellite CW interference. The L1 beamforming trigger follows this first-stage RFI filter.

\textbf{The L1 Trigger: 2-Phi Sector Delay-and-Sum Beamforming: }
The RFSoC SURF will digitize the 8 input channels for each polarization from adjacent azimuthal phi-sectors into the high-performance FPGA fabric. 
A digital low-pass filter at 700~MHz is first applied to the trigger data, which maximizes the impulsive SNR for both Askaryan and air-shower signals due to the higher effective aperture across the payload in this sub-band.
For each polarization, the integrated FPGA will then delay-and-add the 8 channels
together according to plane-wave hypotheses to form 60 synthetic antenna
beams (covering 30$^{\circ}$ in azimuth and 45$^{\circ}$ in elevation)
with gains as high as $16\,\text{dBi}$, sum the squared power received
in every beam, and trigger according to a per-beam impulsive-power threshold. 
Each L1 trigger beam has a -1dB width of 
$\sim$20$^{\circ}$ in azimuth and 3$^{\circ}$ in elevation. 

The L1 beamforming will be run at a high trigger rate of 100~kHz, in order to
maintain the low threshold in the later stages of the trigger.  Event snapshots
for each L1 trigger are shuffled over 20~Gbps direct RFSoC-to-RFSoC links
between neighboring SURF boards to complete the 4-phi sector beamforming as
depicted in Fig.~\ref{fig:pueo_daq}. The L1 event snapshot is a reduced-precision 400~ns long
trace of all 8~channels with 8-bits and 4~GSPS. 

\begin{figure}[tb]%
\centering
\includegraphics[width=0.54\textwidth]{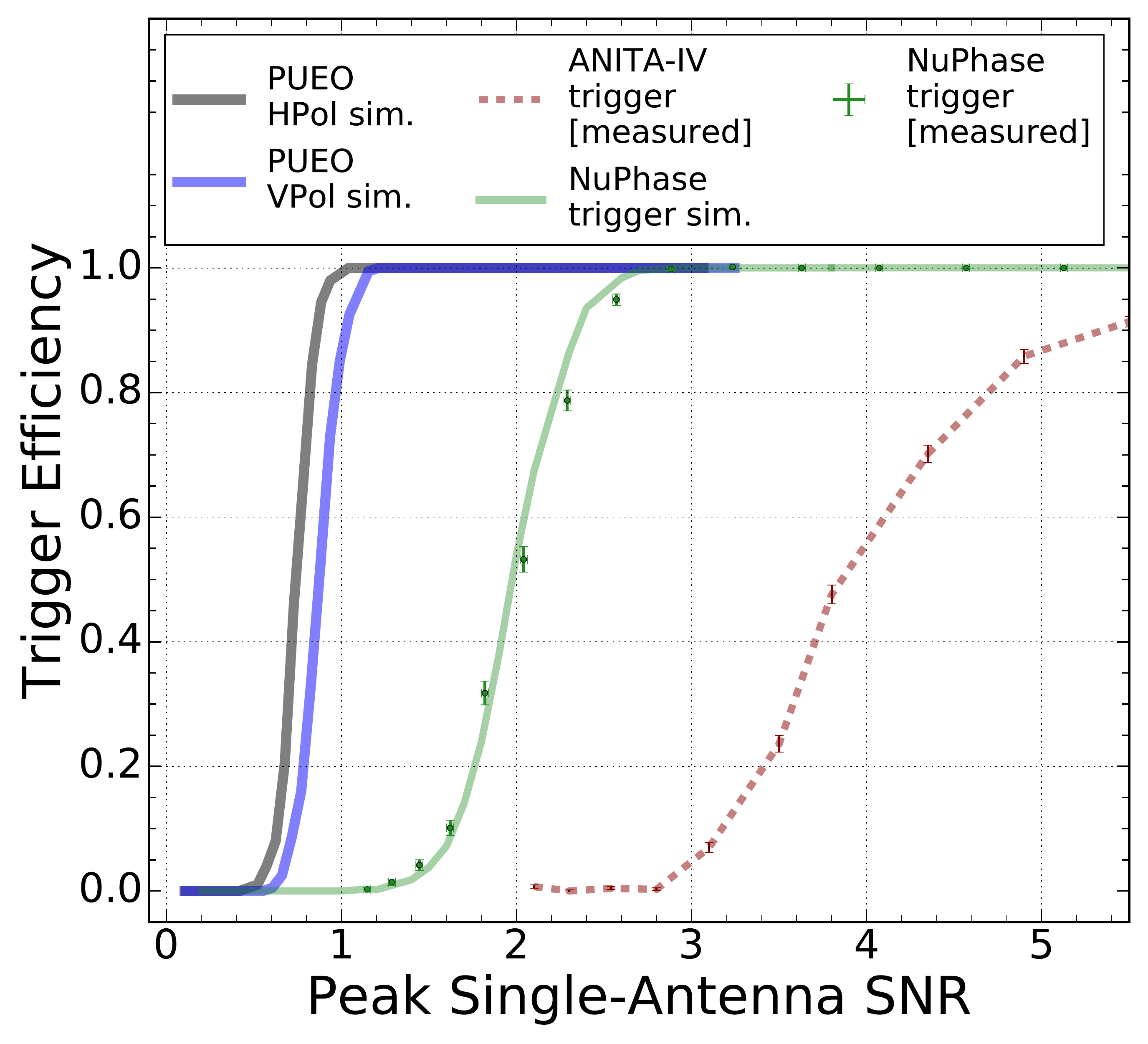}
\caption{A simulation of the 16-antenna PUEO delay-and-sum trigger 
shows a 50\% threshold 
at a voltage signal-to-noise ratio (SNR) of 0.8 as viewed in a single vertically-polarized antenna.
Also shown are the performance of the ANITA-IV combinatoric 
trigger and ARA's coherent trigger system~\cite{oberla}.}
\label{fig:pueo_trigger}
\end{figure}

\textbf{The L2 Trigger, Full 4-Phi Sector Beamforming: }

After an L1~trigger, each RFSoC has collected the full 4-azimuthal phi sector
data from its neighbor SURF and can calculate the full 4-phi sector trigger. It
is likely that the L2 beamforming can be informed by the L1 trigger metadata,
though that has not been fully studied so our simulations assume that each L2
trigger will reform all the beams. The L2 beamformer includes 120 synthetic
antenna beams, which now cover 60$^{\circ}$ in azimuth and maintain the same
45$^{\circ}$ elevation angle coverage. Each adjacent SURF forms a set of 4-phi
sector beams, which ensures that there are some overlapping beams and no gaps
in the trigger coverage.

A full simulation of this 2-stage time-domain beamforming trigger strategy is shown in 
Fig.~\ref{fig:pueo_trigger}, which includes effects from the horn antenna beam 
patterns and impulse response.  We find a 50\% threshold at a voltage SNR
of $0.8\sigma$ (where $\sigma$ is the RMS of the thermal noise), 
for vertically-polarized signals, which is a factor of 5 better than the ANITA-IV trigger performance. 
The horizontally-polarized trigger threshold is further improved (50\% threshold at a voltage SNR
of $0.7\sigma$) due to the wider E-plane horn antenna response in Hpol.
A comparison to the demonstrated 7-channel beamforming 
trigger on ARA~\cite{oberla} performance is also shown, 
which agrees well with a similarly detailed simulation study.  

ANITA has demonstrated a $>99\%$ analysis efficiency, even for near-threshold
events, in isolating signal-like events from thermal noise
backgrounds~\cite{anita4}.  Initial studies indicate that we will be able to
maintain high analysis efficiency for reconstruction of events near the lower
trigger threshold achieved with PUEO, since the SNR of the coherently summed
and deconvolved waveforms used in analysis will be $>5\sigma$.

The anticipated L2 trigger rate under a normal noise environment is expected to be of order 200 Hz. 

\textbf{The L3 Trigger:} At each RFSoC board, the full 16-channel synthetic beam is further processed
to eliminate any CW interference influence on the trigger, including a check on the polarization.
Under normal circumstances without excessive interference, this polarization check will reduce the rate of thermal noise by approximately a factor of two.
If the remaining power is above threshold, 
the region of interest within in the RFSoC recording buffer will be written to disk through the TURF board.
The TURF design for PUEO consists of a
mid-performance FPGA with a 100~Mbps Ethernet link to each RFSoC board, which can handle an overall event-rate-to-disk of $\sim$100~Hz. The TURF also includes
a number of direct FPGA-to-FPGA control lines and distributes a low-jitter clock to each SURF. Two 10-gigabit Ethernet ($10\,\text{GbE}$) links to the
flight computer are included for redundancy, resulting in an overall throughput of $\sim1000\,\text{MB/s}$, much higher than will be practically used.

\textbf{TURF to Flight Computer:}
Once a full payload event is received at the TURF-v5, the data 
is sent to the flight computer via the
dual $10\,\text{GbE}$ links. Though only a subset of digitizer bits are used when forming the trigger, all 12-bits per channel are recorded once the trigger is formed. Each raw PUEO event is  $\sim500\,\text{kB}$, which, at sustained event rate of $100\,\text{Hz}$,
results in $\sim50\,\text{MB/s}$ of data-rate to disk
at the flight computer, the data is directly received into a commercial
general-purpose graphics processing unit (GPU). 
The GPU performs a similar computation as done in the L2 stage of the trigger in order to form a combined directional/signal 
strength metric to generate an event priority for telemetry.

\subsection{The Low-Frequency Instrument}\label{sec:lfsystem}

PUEO will additionally include a low-frequency dropdown instrument. The low-frequency system enhances PUEO's sensitivity to air showers generated by either cosmic rays or tau lepton decays. It additionally serves as research and development (R\&D) aimed at measuring the low-frequency backgrounds across Antarctica over a broad frequency range (50-300 MHz) and independently triggered, which can determine the sensitivity of a future instrument targeting the tau channel specifically. 

{\bf Deployable LF Antennas}
Building on experience from the ANITA-III mission flying the large deployable ANITA Low-Frequency Antenna (ALFA), the PUEO low frequency system is planned to include 4 dual-polarized antennas. The design shown in Fig.~\ref{fig:rendering} includes a drop-down consisting of crossed, electrically-short bowtie antennas. However, other antennas will be considered through a future trade study. The LF system will be deployed after launch, having been stowed on the interior of the payload prior to launch. %

{\bf Dedicated Trigger for the Low-Frequency (LF) Instrument: }
The 8-channel low-frequency instrument will also have a two-stage beamforming trigger
similar to the main payload trigger, but computed on a single LF SURF board.
The full beamforming solution can be calculated in a
single stage on the LF SURF; the integrated power of the peak beam is then compared against a
per-beam impulsive power threshold to decide on an L1-trigger. The synthetic
waveform is then further processed to eliminate CW interference as
well as spectral comparisons against known properties of EAS events before
deciding on an L3 trigger. A full simulation of the LF instrument for
EAS events finds a 50\% trigger SNR of $\sim1.8\sigma$ at a single LF low-gain antenna.

An L3 trigger in the LF instrument will trigger event data from the entire
payload (LF+Main Instrument) to be stored. Similarly, an L3 trigger in the main payload will
also save event data from the LF instrument. To further reduce our threshold,
the synthetic beams from L2 triggers of the Main Instrument and LF instrument are further
combined to check for an L3 trigger even if neither instrument triggers independently.
The low-frequency instrument trigger is expected to contribute negligibly ($\mathcal{O}$(1 Hz)) to the total trigger rate. 

\begin{figure}[tb] 
  \centering
  \includegraphics[height=0.42\textwidth]{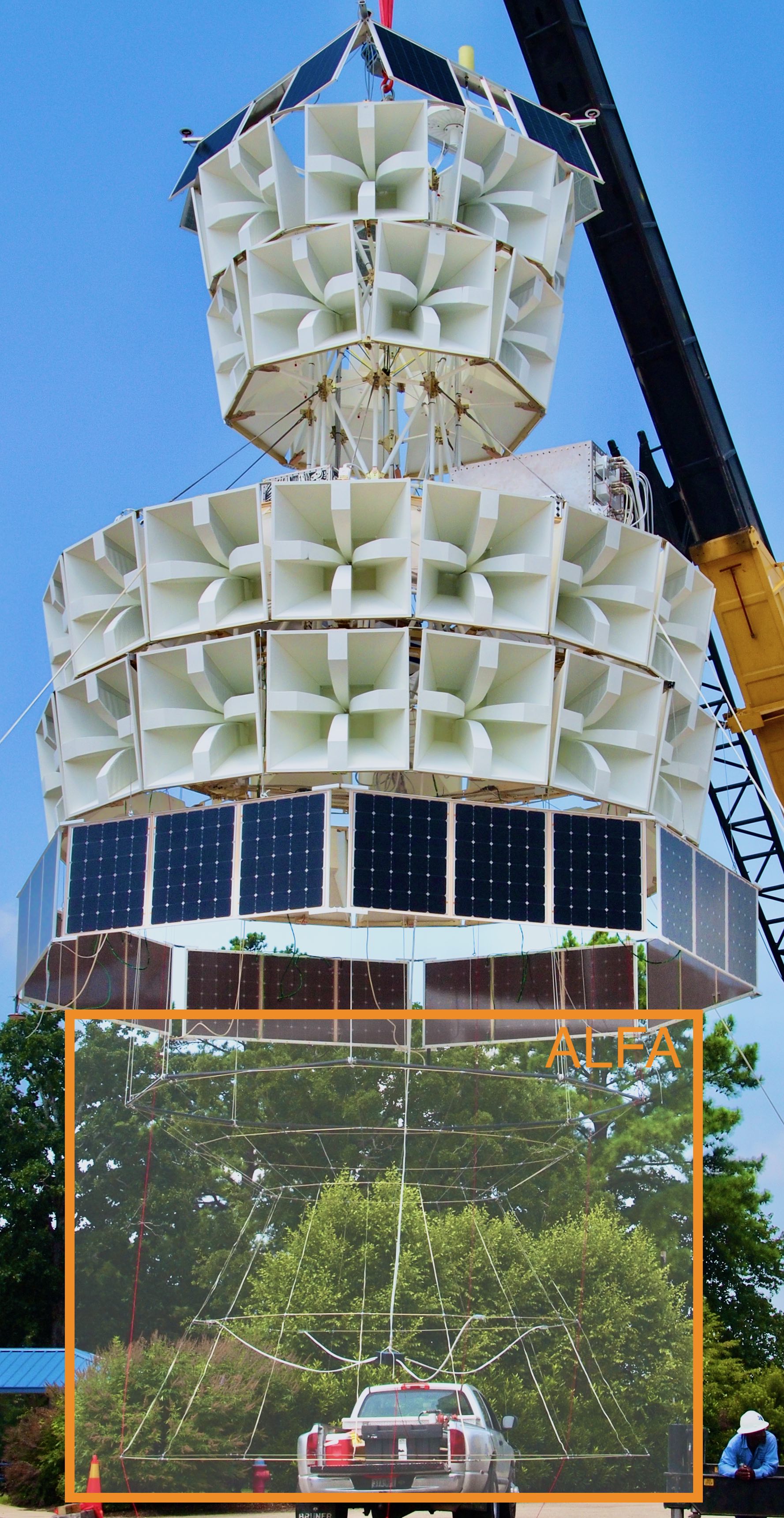}
  \includegraphics[height=0.42\textwidth]{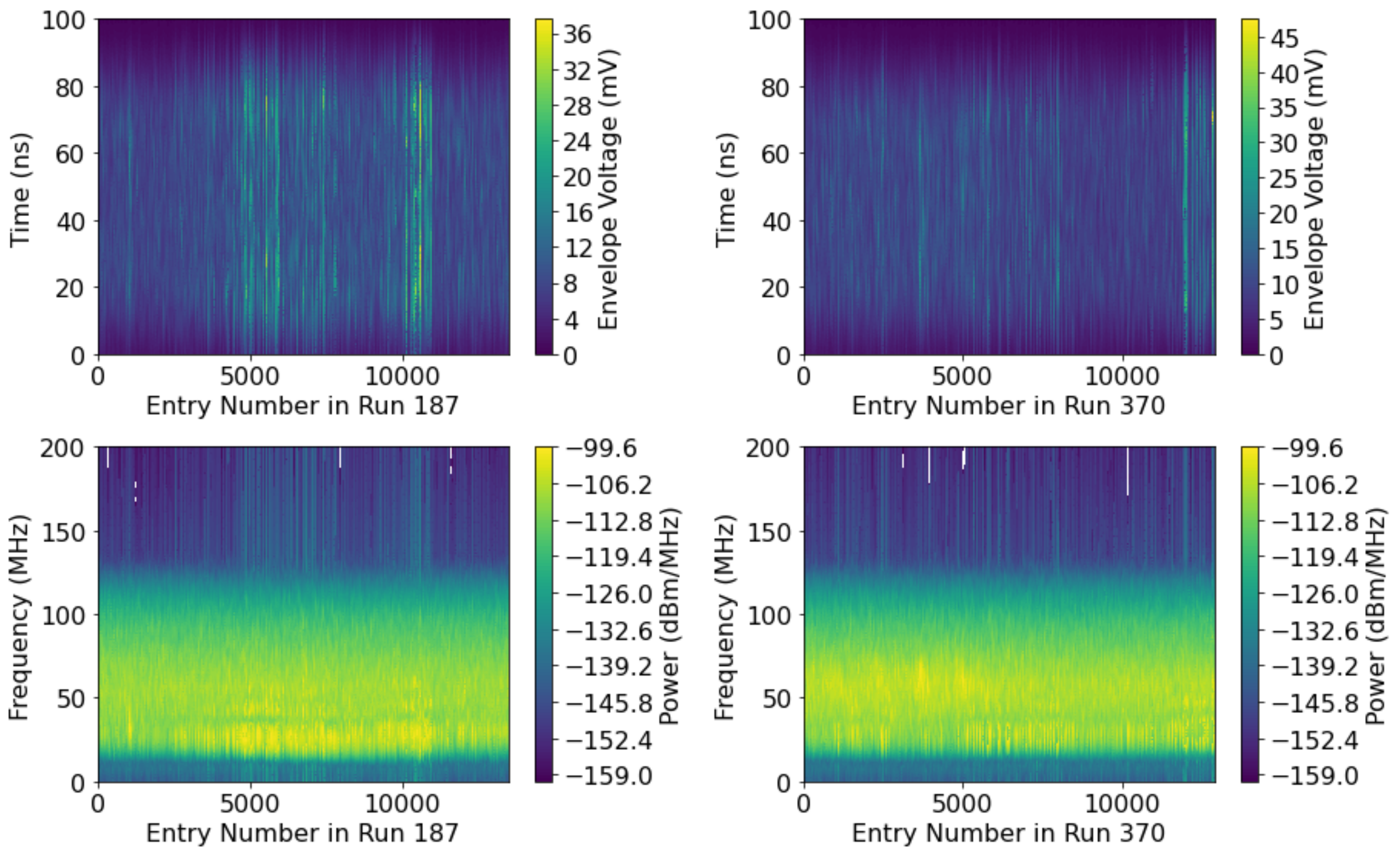}
  \caption{The ALFA antenna on ANITA-III. The left panel shows the fully deployed ALFA antenna below ANITA-III during the hang test at the Columbia Scientific Balloon Facility. The middle and right panels show minimum-bias data from the ANITA-III ALFA instrument at two times during the flight (middle panel: East Antarctica and right panel: West Antarctica). The top panels shows the maximum amplitude of ALFA voltage envelope while the bottom shows a spectrogram. During the flight the waveforms from the ALFA were upconverted to the frequency band of the ANITA DAQ; these waveforms and spectra are downconverted to the ALFA baseband of 40-80 MHz.}
  \label{fig:alfa} 
\end{figure}

{\bf ANITA heritage:} The ALFA was a prototype
drop-down low-frequency instrument that was part of the ANITA-III flight. ALFA
was a single antenna of a custom design based on a deployable, tapered quad
slot that formed an omni-directional horizontally-polarized antenna. While it
did not trigger ANITA-III, ALFA was able to measure the low-frequency spectrum
of ANITA-III events~\cite{anita3me}. Minimum-bias data from ALFA
(Fig.~\ref{fig:alfa}) suggests that the lower-frequency bands are clean much of
the time in Antarctica, allowing for the successful use of the LF instrument.

\subsection{Data Storage and Transmission}

A 30-day flight of the PUEO instrument will generate $\sim50\,\mbox{TB}$ of data, including data compression.
Previous flight experience with ANITA has stressed the importance of triply-redundant
storage and multiple storage technologies to guarantee a complete
copy of the in-flight dataset. PUEO will use storage technologies
previously proven on both ANITA-III and~-IV, consisting of a
primary storage system directly connected to the flight computer
and a secondary storage computer connected via Ethernet. The primary
storage system consists of two arrays of four commercial sealed helium-filled
spinning magnetic drives of at least $20\,\mbox{TB}$, for a total
primary storage of $2\times80\,\mbox{TB}$. The secondary storage
computer from previous flights will be upgraded with eight solid-state
storage devices of at least $10\,\mbox{TB}$ each, for a total secondary
storage of $80\,\mbox{TB}$. 
In both the primary and secondary
storage systems, only one of the drives in an array will be active
at one time, both to minimize power and risk. The storage systems
will be located at an easy-to-access panel at the instrument enclosure
to guarantee data recovery, even when the primary instrument may not
be immediately recoverable.

During Line-of-Sight (LOS) operations, PUEO will use a 1 Mbps S-band downlink for parameter tuning, coupled with slow-speed Iridium data modems and a P-band command uplink. This biphase data will be received and decoded locally in McMurdo. %
Once out of the LOS period, PUEO will switch to downlink via the Telemetry Data Relay Satellite System (TDRSS), and the prioritized data will be received and distributed from a dedicated server at the ground station to several PUEO institutions. TDRSS supports both 6 kbps and 90 kbps modes.  The physics trigger rates on PUEO are low enough that the high-priority  data captures the highest-quality physics events. Telemetered data will be used for flight monitoring and serve as an additional contingency in the rare case that the the triply-redundant storage system fails. 
For command and control, an Iridium data modem provided by the NASA ballooning program, will be used. The command system will also be available via low-rate TDRSS as a backup.

\subsection{Gondola}

The gondola design is based on the ANITA gondola structure, benefiting from ANITA’s proven flight heritage. The base structure comprises carbon fiber tubes fitted with 7075 aluminum alloy (AA7075) ends and custom extruded tee sections. Components are assembled using titanium pins for rapid field disassembly. Additional light-weighting and magnesium alloy substitutions will be employed, where applicable, to keep mass inline with balloon capacity. PUEO’s scaled down ANITA quad ridge horn antennas (QRH) will allow for an additional 48 antennas on the main gondola while maintaining adherence to the prescribed launch envelope and mass budget. PUEO will also include a $-40^\circ$ canted 12 QRH antenna nadir ring and LF antenna array both of which will be stowed up and within the lower gondola section in a pre-launch configuration then deployed by command during the first moments post-launch. The photovoltaic (PV) panel array will also be stowed in a raised position during pre-flight then deployed post-launch. Deployable systems, utilized in the past three ANITA flights, allow for the expansion of at-float payload capabilities while maintaining adherence to the launch envelope on the ground. The current best estimate (CBE) of the science-only payload mass is 3,618\,lbs. The CBE summed mass of ballast, science, and NASA systems below the gondola suspension point is 4,331\,lbs. 

\subsection{Power System}

The PUEO power system will be derived from the ANITA-III and~-IV systems. 
We will use an icositetragon (24-gon) omnidirectional array of solar panels. The ANITA ``skirt''-array 
design has been found to be robust, with the vertical photovoltaic (PV)
angles enabling a substantial power contribution from sunlight
reflected off of the ice, in additional to direct sun. 
This power can be provided using 140-cell strings 
(SunPower E-60 solar cells; 23\% efficiency) on each side. 
The panel size can be increased by $\sim$71\% compared to ANITA-IV, 
giving a maximum PUEO PV power envelope of 1,625\,W.
The cells will be laminated onto dielectric honeycomb substrates, as was done 
successfully with ANITA-III and~-IV. The array will be lowered immediately after launch, 
using a similar mechanism to that in ANITA-III and~-IV.
We will use a Morningstar Tristar-MPPT-60 charge controller, as was used on ANITA-III and~-IV. 
The PUEO battery system will consist of 6 pairs of Panasonic LC-X1220 Pb-acid batteries 
(20 Ah each). %

The power distribution system consists of three main power rails:
a main instrument box rail, nominally $12\,\text{V}$ with a load
capability of $500\,\text{W}$, and two low-voltage RF signal-chain rails at
$4\,\text{V}$ (for pre-amplifiers external to the instrument box) and
$3.7\,\text{V}$ (for internal RF components) with
load capabilities of $200\,\text{W}$ each. Local board-level voltages
at the instrument box are generated via high efficiency ($95\%$)
point-of-load regulation. Including  DC/DC  conversion efficiencies for the PV system,
this brings the total payload power CBE to 1,139\,W, leaving ample margin.

\subsection{Attitude and Location Determination}

PUEO has requirements for pointing resolution in elevation that are more
stringent than ANITA for two reasons.  First, better pointing resolution in
elevation allows PUEO to better distinguish above-horizon from below-horizon
events.  This is important for detection of possible EASs in the tau neutrino
channel, which for Standard Model cross sections would tend to come from near
the horizon. Second, a better uncertainty on elevation angle leads to better
rejection of man-made events while maintaining high analysis efficiency, since
man-made noise can be better localized when projected onto the continent,
especially near the horizon. This leads to a requirement of 0.05$^{\circ}$
accuracy in pitch and roll.  An $0.05^\circ$ heading accuracy will ensure that
the natural rotation of the payload does not restrict event reconstruction in
azimuth.

Similar to ANITA, PUEO will use a multiply-redundant system to determine
absolute orientation.  PUEO will incorporate a Trimble ABX-TWO, which provides
better angular resolution than the ADU5, which was flown on ANITA, in
high-latitude regions.  

In
addition, PUEO will incorporate a Northrop LN-251, which is an inertial
measurement unit (IMU), integrated into an RF enclosure.  This will provide
precision pitch and roll information, required for PUEO's more stringent
pointing requirements compared to ANITA.  The LN-251 is integrated into the
Wallops Arc Second Pointer (WASP) system and has been successfully used in
flight in Antarctica~\cite{wasp,xcalibur}.

PUEO will additionally incorporate two daytime star trackers to perform real-time
attitude reconstruction using bright-star astrometry. Daytime star trackers
have flown on previous Antarctic LDB
payloads~\cite{2006:Rex_BLAST,2014:Chapman_EBEX} and provided $\sim4$"
($0.001^\circ$) accuracy at~$\sim1$~Hz over the duration of the previous
flights. This flight-proven accuracy exceeds the attitude requirements by more
than an order of magnitude. Preliminary Monte Carlo simulations and ground
measurements using the expected flight hardware have replicated these results
for the PUEO tracker design.

PUEO will also have a magnetometer and sun-sensor systems.  The flight magnetometer is expected to deliver $0.1^\circ$ accuracy orientation when combined with the sun-sensor data.  These constitute a credible back up system that would not degrade performance significantly. %

\subsection{Calibration of the PUEO Instrument}

PUEO will employ a multi-pronged approach to instrument calibration that builds on the calibration systems used on ANITA. 

\subsubsection{Lab Measurements}

Measurements of the gain, noise figure, and trigger efficiency will be made
prior to flight and will be used to benchmark the performance of the
instrument. Lab tests will use broadband, neutrino-like pulses as well as
cosmic-ray-like pulses with more low-frequency content injected into the front
end amplifiers to measure the trigger efficiency curve shown in
Fig.~\ref{fig:pueo_trigger}.

\subsubsection{Ground-to-Payload Calibration Systems}

Ground-to-payload radio calibration pulses are required for the analysis of
PUEO data and for in-flight instrument verification. For all ANITA flights, we
transmitted broadband, high-voltage neutrino-like pulses from the ground at
known positions and rates tied to the GPS-second.

These pulses are used for a variety of purposes, including: (1) determination
of the in-flight antenna phase centers; (2) trigger and analysis efficiency
in-flight as a function of SNR; and (3) measuring the pointing accuracy.

For PUEO, we plan to transmit calibration signals with varying frequency content and pulse shape using a programmable pulser in addition to the signals from the high-voltage pulsers that were used to calibrate ANITA. This will allow us to further refine the calibration of antenna phase center locations for PUEO.
Ground calibration stations will be located near the launch site at LDB and in remote locations as they were in prior ANITA flights. Directly after launch the station at LDB allows for immediate verification of system performance.
The calibration stations at deep field sites are used to acquire science-level calibration data because they are less contaminated by background interference. 
For the PUEO remote stations, we propose to establish two manned stations at established camps such as WAIS Divide or Siple Dome. The redundancy enhances the likelihood that the PUEO payload will fly within the 700~km horizon of one of the two camps. Both proposed stations have been along the flight paths of both ANITA-III and~-IV.
Manned calibration stations at these or similar locations have ensured the full calibration of previous ANITA flights.

\subsubsection{HiCal}

The third iteration of the High-altitude Calibration payload (HiCal-3) will be a key component in the PUEO calibration suite. Building upon the previous two flights~\cite{hical1, hical2, hical_instrument}, HiCal-3 is planned to comprise a set of three payloads, with extended capabilities for pulsing, capturing, and transmitting data relative to previous flights. In addition to providing a regular calibration for PUEO, HiCal-3 will provide crucial data for studying various properties of radio reflection from the Antarctic surface. This is of heightened importance in view of the recently reported anomalous events~\cite{anita3me}. Here we describe the instruments and the role they play in PUEO's science case.

HiCal is a calibration payload with GPS location and satellite telemetry capability that flies in tandem with PUEO, broadcasting calibration signals at regular intervals that are received directly and also reflected from the surface of the continent. Because the PUEO and HiCal payloads are launched separately, the distance between them varies throughout the flight. This allows the surface of the continent to be probed over a range of incidence angles. However, because the flight paths of high-altitude balloons are uncertain, it is generally prudent to launch multiple payloads, to ensure proximity to PUEO for as much of the flight as possible. HiCal-2 comprised 2 payloads, both of which were (fortunately) in proximity with ANITA-4 for the duration of their flights, but this is not guaranteed. To improve the chance of success, we will launch a suite of three payloads, one `primary' payload, and a pair of smaller, `secondary' payloads, with less functionality but similar calibration capability. Of the three, the primary payload is the only one for which we anticipate recovery.
\begin{figure}[t]
\centering
  \includegraphics[width=.7\textwidth]{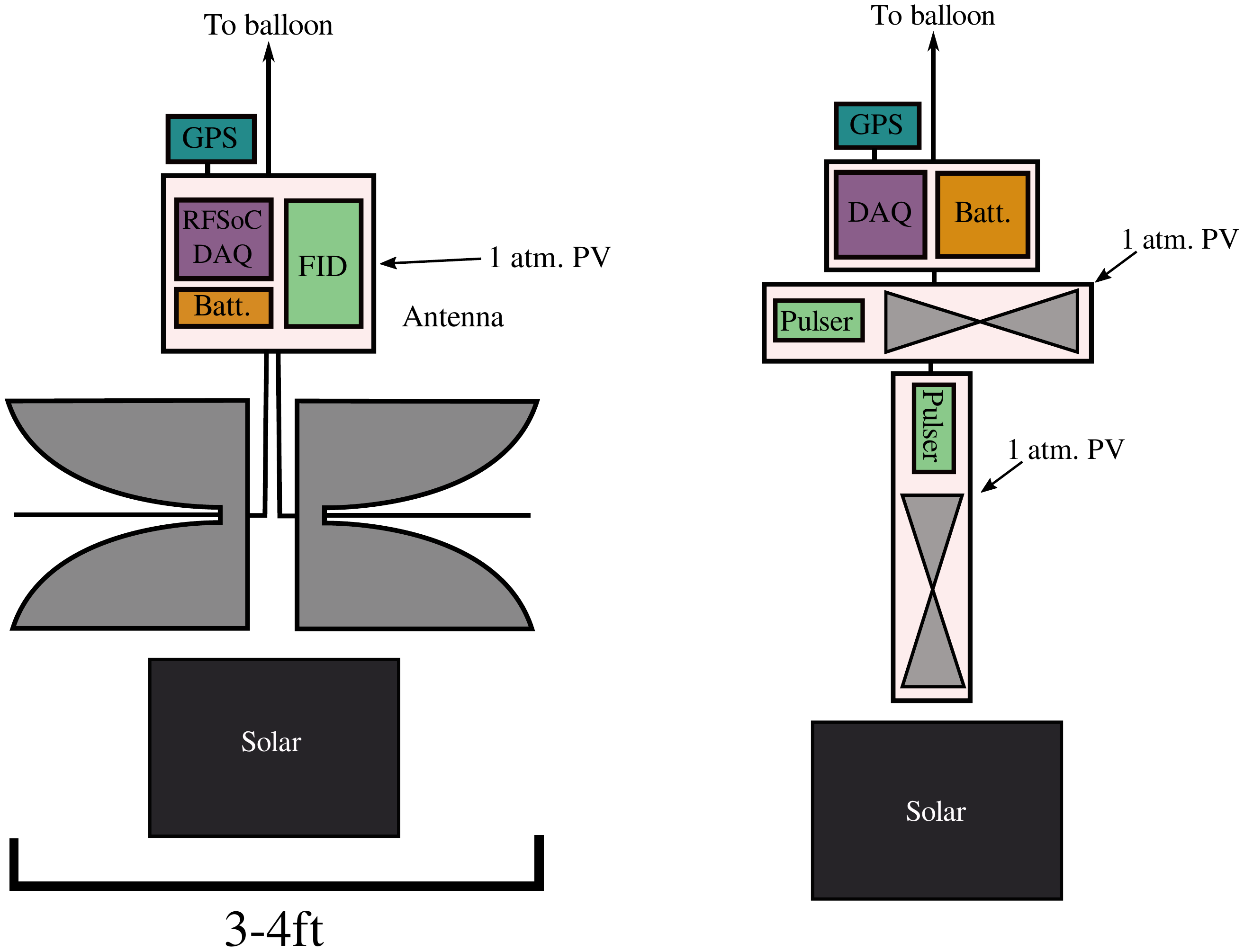}
  \caption{The primary (Left) and secondary (Right) HiCal-3 payloads. Indicated are all subsystems; pressure vessels facilitate the use of high-voltage in the upper atmosphere. The primary payload has two dual-polarization Vivaldi-style antennas, potted to protect against high-voltage arcing. The secondary payload has 2 individual pressure vessels, one for horizontal polarization, and the second for vertical polarization. Not to scale.}
  \label{fig:hc3}
\end{figure}

The primary payload (Figure~\ref{fig:hc3}, left) will be solar-powered, and have a low-throughput satellite link, local disk storage, and on-board signal generation and digitization capabilities. It will include a set of dual-polarized, wide-band antennas and a high-voltage commercial FID pulser to produce sharp impulsive signals, probing reflectivity throughout the PUEO band. Also on board will be an RFSoC ADC/DAC providing the capability to produce arbitrary waveform radio signals to further calibrate the PUEO instrument and to timestamp and digitize signals from a local receiving antenna. We also plan to locally digitize and capture the GPS signals themselves, as these signals can be used to model RF propagation in the atmosphere, which is particularly challenging near the horizon, and holds significant interest for earth-skimming events (several of which have been detected by ANITA). As with HiCal-2, we anticipate that azimuthal payload orientation will be monitored using an array of silicon photomultipliers sensitive to solar emissions. The secondary payloads (Figure~\ref{fig:hc3}, Right) will be slimmed-down versions, with custom solid-state pulser units built in-house at KU (already demonstrated for the TAROGE-M~\cite{taroge-m} experiment as well as broadcasts to the ARA experiment at South Pole from within the SPICE core~\cite{spicecore}), low-throughput satellite telemetry, and no on-board digitization. These secondary units would therefore accomplish the main goal of calibrating PUEO, but would not offer the same breadth of signal shape and internal cross-checks. Lacking the significantly more expensive commercial hardware of the primary payload, we do not anticipate recovery of the secondary payloads. 

\subsubsection{Passive Calibration} 

In addition to the active calibrations listed above, the calibration of the
antenna positions can be checked with passive measurements of the sun and
satellites. In particular, signals from  low-band (1.2 GHz)  GNSS satellites
will be measured in every PUEO event at levels below thermal noise but
reconstructible using interferometric analysis. The reconstructed positions of
the satellites can be compared to the well-known ephemerides in analysis to
check the stability of pointing precision on an event-by-event basis. 

\section{PUEO Expected Performance} 

The upgrades listed in the previous sections will significantly reduce the
trigger threshold of PUEO compared to ANITA. As can be read off
Fig.~\ref{fig:pueo_trigger}, the phased-array trigger with the RFSoC will
result in an electric-field threshold reduction factor of at least a factor of
5 in the Askaryan channel and 7 in the EAS channel using the high-frequency
instrument.  In this section, we show how this lower threshold results in
substantial sensitivity improvements compared to ANITA and other instruments.
For example, at 30 EeV, PUEO is expected to be approximately two orders of
magnitude more sensitive than ANITA to a diffuse flux.  The low-frequency
instrument will also provide improved sensitivity to neutrinos through air
showers well compared to ANITA. While the Askaryan channel still dominates over
the air shower channel  except at the lowest energies, the air shower channel
has improved neutrino direcctional reconstruction and, combined with the
Askaryan channel, offers some possibility of flavor identification.

\subsection{Diffuse UHE $\nu$ Sensitivity} 

\begin{figure} 
  \centering
\includegraphics[width=0.9\textwidth]{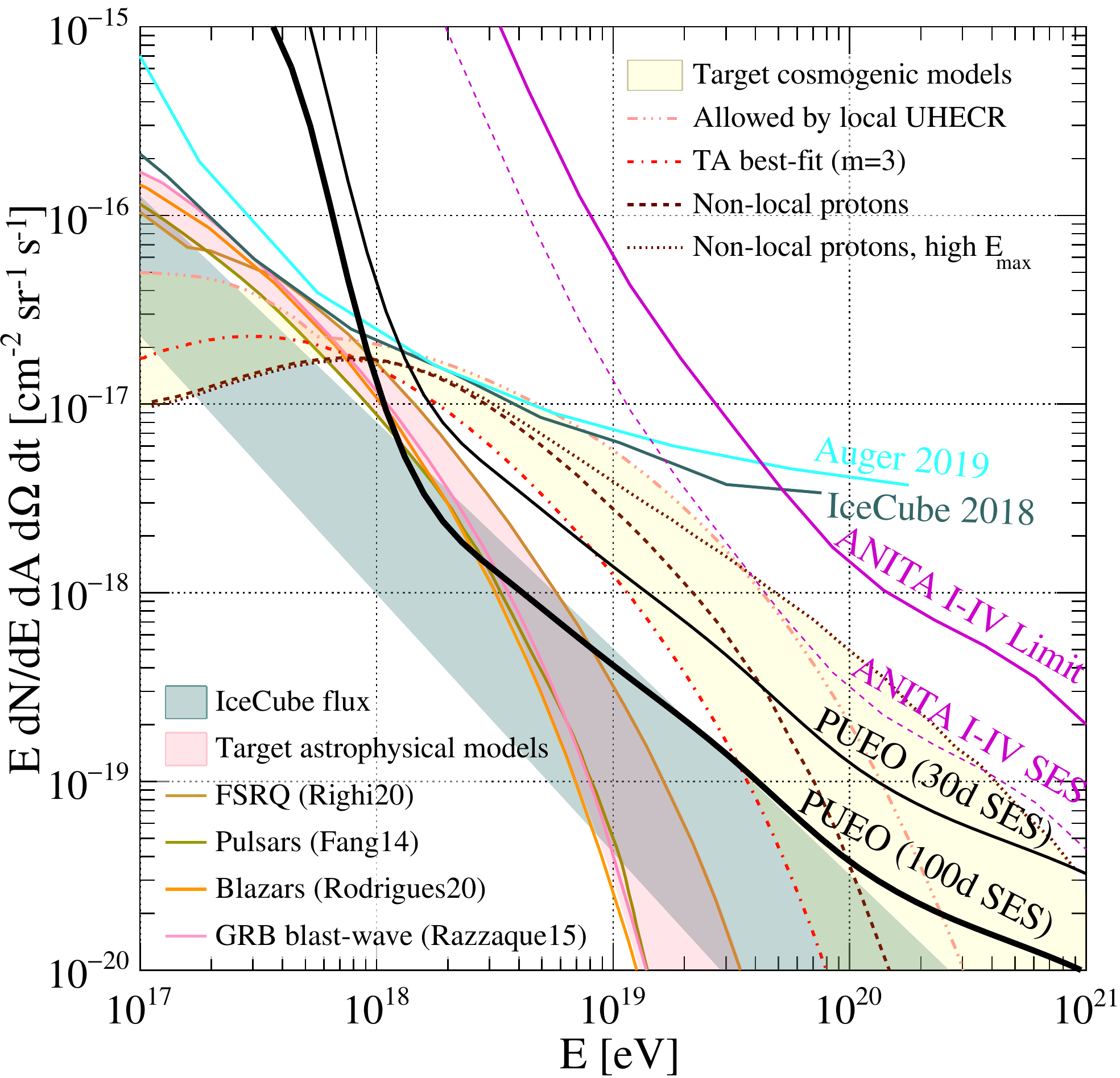}
  \caption{The PUEO single-event sensitivity (SES) to diffuse UHE fluxes, compared to
  existing limits ~\cite{auger2019, Aartsen:2018vtx, anita4} and some
  cosmogenic models~\cite{vanVliet:2019nse,TA_composition_1, TA_composition}
  and astrophysical models ~\cite{righi2020eev,FangPulsar,rodrigues2020blazar,GRBBlastwave}. The ANITA I-IV SES is shown for comparison. The non-local proton models
  were generated using CRPropa3 in a manner similar to ~\cite{vanVliet:2019nse}
  but with $z>0.1$. For diffuse fluxes, the Askaryan sensitivity dominates, although the $\tau$ EAS channel also contributes significantly below a few EeV.}
\label{fig:diffuse_sensitivity} 
\end{figure}

Tools developed for ANITA were adapted to estimate the projected sensitivity of PUEO to a diffuse flux of UHE
neutrinos in both the Askaryan and $\nu_\tau$ air shower channel. 

For the Askaryan channel, the \texttt{icemc}~\cite{icemcPaper} simulation package 
was adapted to compute the Askaryan acceptance for PUEO by introducing the
appropriate threshold-scaling compared to ANITA-IV. As the total exposure depends on the flight
trajectory and duration, the ANITA-IV flight path was used
and the exposure scaled to a typical flight time of 30 days. A small correction
factor was made to the effective area for the improved RFI filtering that will
be possible with PUEO. As with previous results from ANITA, the effective areas
are geometrically averaged with a projection from an independent simulation,
which includes the effects of surface roughness which are currently not modeled
in \texttt{icemc}.

The $\nu_\tau$ air shower channel is modeled with a dedicated simulation based
on ~\cite{anita_tauexposure}, with appropriate threshold and noise scaling for
the high-frequency and low-frequency bands. Above 1 EeV, the Askaryan channel
dominates in effective area for a diffuse flux. 

The trigger-level single-event sensitivity (SES) is computed at fixed energies using: 

\begin{equation}
  \mathrm{SES}(E) = \frac{1} { T \cdot \Delta \cdot A\Omega_{\rm{eff}}(E)}, 
\end{equation}

\noindent where $T$ is the flight duration,
$\Delta=4$ is a bandwidth factor (see below) as in previous ANITA results, and
$A\Omega_{\rm{eff}}$ is the aperture (area $\times$ solid angle) of PUEO at the energy.  The SES is compared to
existing experiments and some astrophysical and cosmogenic models in
Fig.~\ref{fig:diffuse_sensitivity}, where the SES has been splined for
visualization purposes. We find that a 30-day flight of PUEO will either
measure or likely eliminate a number of cosmogenic models from non-local or
subdominant proton sources. A 100-day campaign can confirm or exclude the
best-fit TA composition and could also measure diffuse astrophysical neutrinos
from FSRQs, Pulsars and GRBs. If the IceCube astrophysical flux extends to PUEO
energies, that may also be observable. 

The sensitivity curve bandwidth ($\Delta$) accounts for the rapid evolution of
instrument acceptance and neutrino flux with energy, and therefore is not 
model-independent. The choice of $\Delta=4$ was identified as a compromise between
different models~\cite{rice06} and we continue to carry forward the same
convention. With this choice, a model exactly equaling the SES for a decade in energy results in $\log(10)/\Delta \approx 0.57$ events.
To compute the number of events expected to be observed for a particular model, the model flux may be integrated the exposure ($TA\Omega_{\rm{eff}}$). For a diffuse flux,

\begin{equation}
  N = T \int_E F(E) A\Omega_{\rm{eff}} dE. 
\end{equation}
Expected triggered event counts for some particular models are
tabulated in Tab.~\ref{tbl:event_counts}.

\begin{table} 
  \centering
\begin{tabular}{l|c|c}
Model & $<N_{\nu}>$  (30 d) & $<N_{\nu}>$  (100 d) \\ 
\hline
\hline
Non-local proton cosmogenics &  1.8 & 5.9 \\ 
Non-local proton cosmogenics (high $E_{max}$) &  4.5 & 15.0 \\ 
TA best fit ~\cite{TA_composition,TA_composition_1}&  0.9 & 2.9 \\ 
Subdominant proton cosmogenics~\cite{vanVliet:2019nse} &  3.7 & 12.4 \\ 
\hline
Diffuse FSRQs, max~\cite{righi2020eev} &  0.4 & 1.3 \\ 
Diffuse Pulsars, max~\cite{FangPulsar}  &  0.2 & 0.5 \\ 
Diffuse AGN~\cite{rodrigues2020blazar} &  0.2 & 0.5 \\ 
Diffuse GRB Blast-waves, max~\cite{GRBBlastwave} &  0.2 & 0.6 \\ 
IceCube flux, max, all-flavor, extrapolated~\cite{IceCubeFlux} &  0.5 & 1.7 \\ 
\hline
\end{tabular}

   \caption{PUEO diffuse UHE neutrino event counts for various models. The top four are different cosmogenic models, while the bottom five are astrophysical models. The total number of observed events can be a combination of a cosmogenic model and one or more astrophysical models. } 
  \label{tbl:event_counts} 
\end{table}

\subsection{Transient UHE $\nu$ Sensitivity} 

\begin{figure} 

  \centering
  \includegraphics[width=0.49\textwidth]{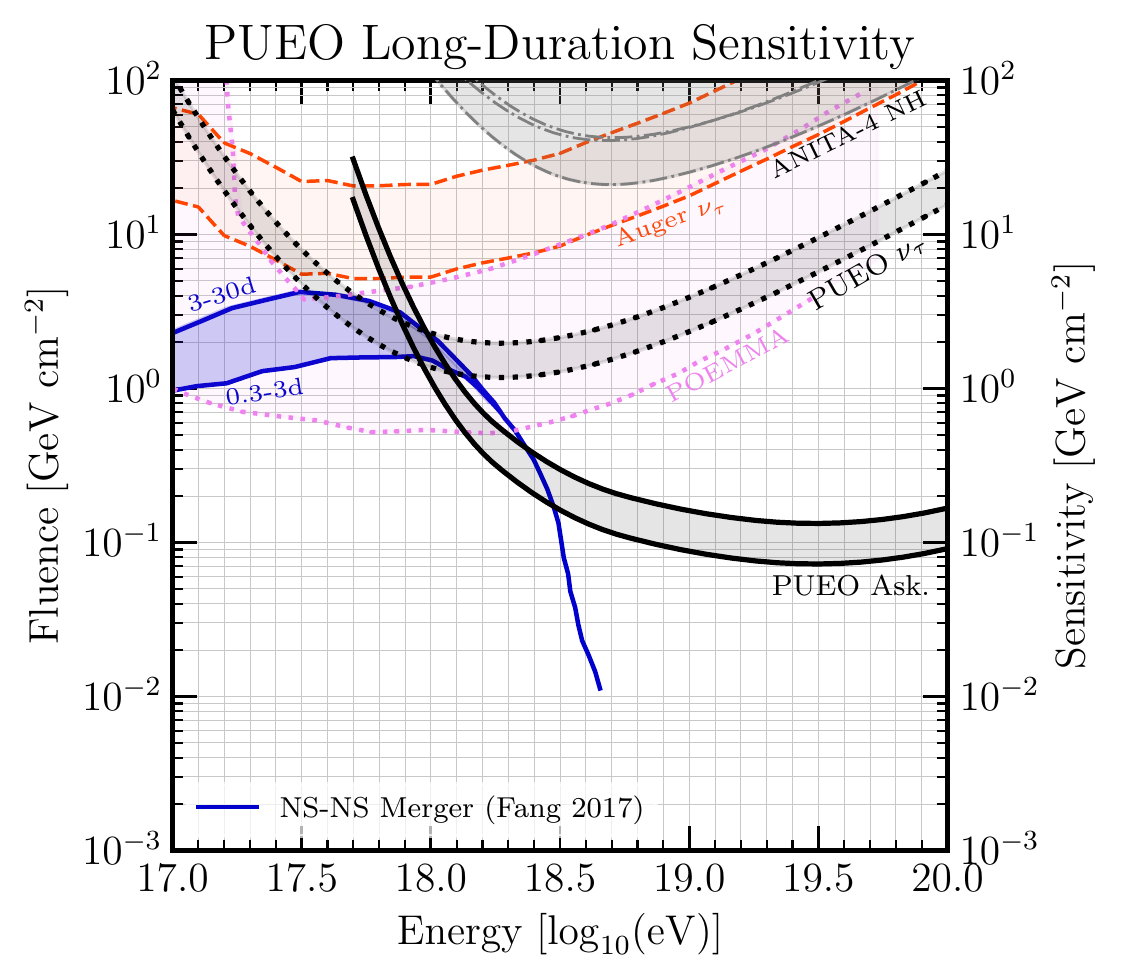}
  \includegraphics[width=0.49\textwidth]{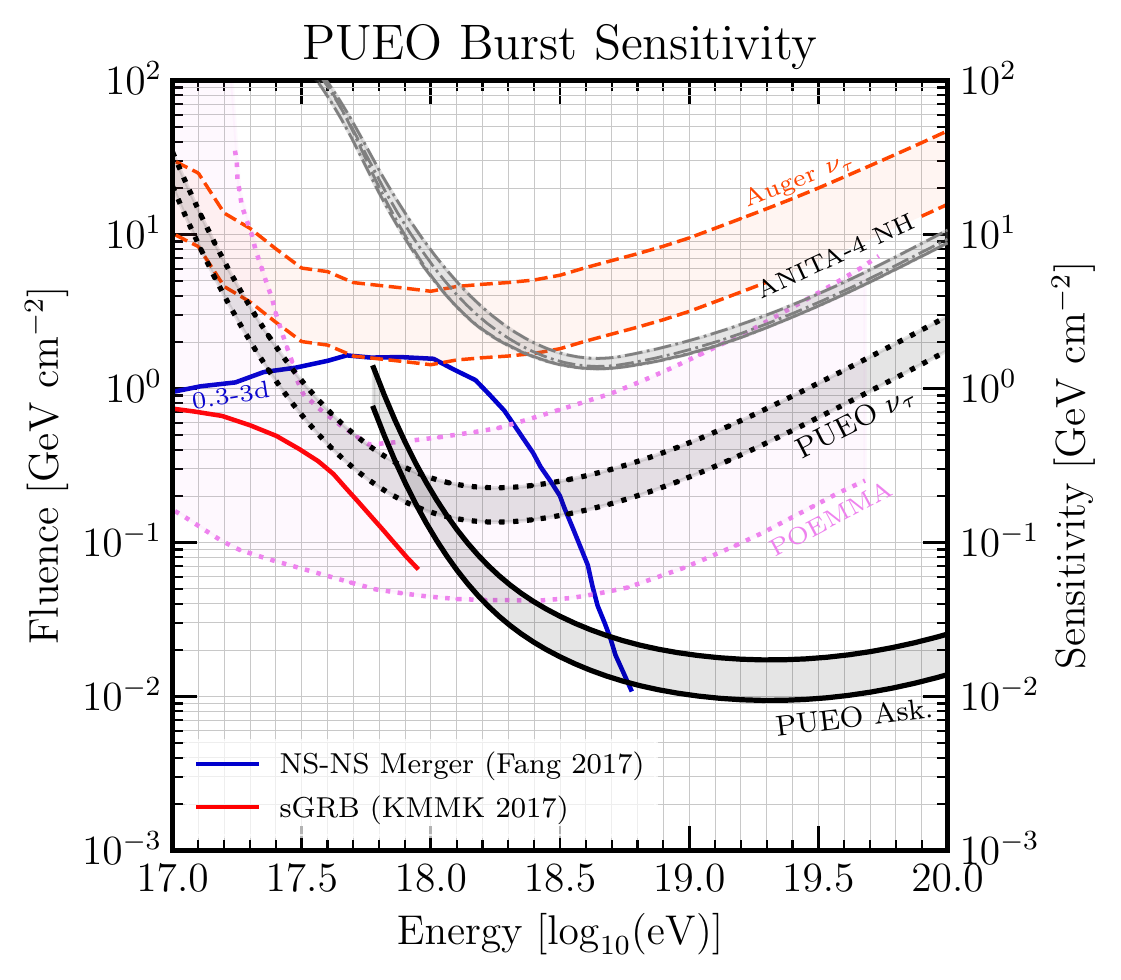}

\caption{The peak single-event sensitivity of the PUEO air shower and Askaryan
  channels to long (left) and short (right) transients. The long transient
  sensitivity considers the mean effective area of an optimal part of the sky
  over the course of the flight, while the short burst sensitivity considers
  the peak effective area over a 1000s window, which is appropriate for
  transients of a few hours or smaller. The bands shown for the various
  experiments denote the range between the optimal and average sky sensitivity,
  for regions where the instrument has sensitivity.} 

\label{fig:transient} 
\end{figure}

PUEO's very large instantaneous aperture makes PUEO well-suited to measuring
UHE neutrino fluence from transient astrophysical sources that occur in its
most sensitive field-of-view. Fig.~\ref{fig:transient} shows PUEO's peak
fluence sensitivity for both ``long" and ``short" transients. Long transients
are of order the length of the flight, so the point in the sky is not being
constantly monitored by PUEO. The average effective area over a typical flight
path for a source at a favorable declination is used to determine the
peak long-duration sensitivity. A short transient is of order a few hours or
less, where the source may remain within PUEO's peak instantaneous sensitivity for the entire duration,
allowing even greater fluence sensitivity for objects in that angular range.

Examples of transient astrophysical phenomena that may produce detectable
fluxes include
neutron-neutron star mergers~\cite{Metzger} and short GRBs~\cite{NSBMH}. Should the recently-announced near-horizon air
showers from ANITA-IV~\cite{anita4cr} be from transient sources of $\tau$
neutrinos, PUEO would likely detect orders of magnitude more of this class of events. PUEO's transient sensitivity also
compares well to the proposed POEMMA satellite mission~\cite{poemma,Venters:2019xwi}.

The Askaryan transient sensitivity was estimated using the PUEO-configured
\texttt{icemc} in point source mode as described in ~\cite{anita3source}, with
an ice-roughness correction applied based on an independent simulation. For the
Askaryan channel, the peak instantaneous sensitivity varies depending on the
position of the payload, with deeper colder ice providing additional
sensitivity. 

The air shower channel sensitivity is estimated with the same dedicated air
shower simulation package used in the diffuse case, which supports both point
source and diffuse sensitivities. The air shower channel has a very narrow peak
sensitivity right at the horizon, which becomes dominant to the Askaryan
channel sensitivity below 1 EeV.

\subsection{Direction and Energy Reconstruction} 

If PUEO is able to make a confident detection of UHE neutrinos, whether diffuse
or associated with a transient, the next step would be to characterize the
properties of the detected neutrinos.  The air-shower channel provides excellent
neutrino pointing resolution (of order a degree) due to the narrow emission cone. In the
Askaryan channel, the opening angle of the in-ice emission cone is large, so
estimating the neutrino direction requires measuring both the apparent 
radio direction and the polarization vector of the electric field.  We expect that the neutrino direction may be
reconstructed to a region (68\% confinement contour) of order 15 square degrees
~\cite{anita3source}.

Energy reconstruction of the neutrino is ultimately limited by the inelasticity
in the primary interaction~\cite{Connolly:2011vc}, although making assumptions about the neutrino energy spectrum can
help. The shower energy can be estimated with the measured electric
field strength as well its power spectrum that encodes the off-cone angle (and,
to some extent for the Askaryan channel, the length of the signal path through the ice).

\subsection{Flavor Physics}

\begin{figure}
  \centering
  \includegraphics[width=4in]{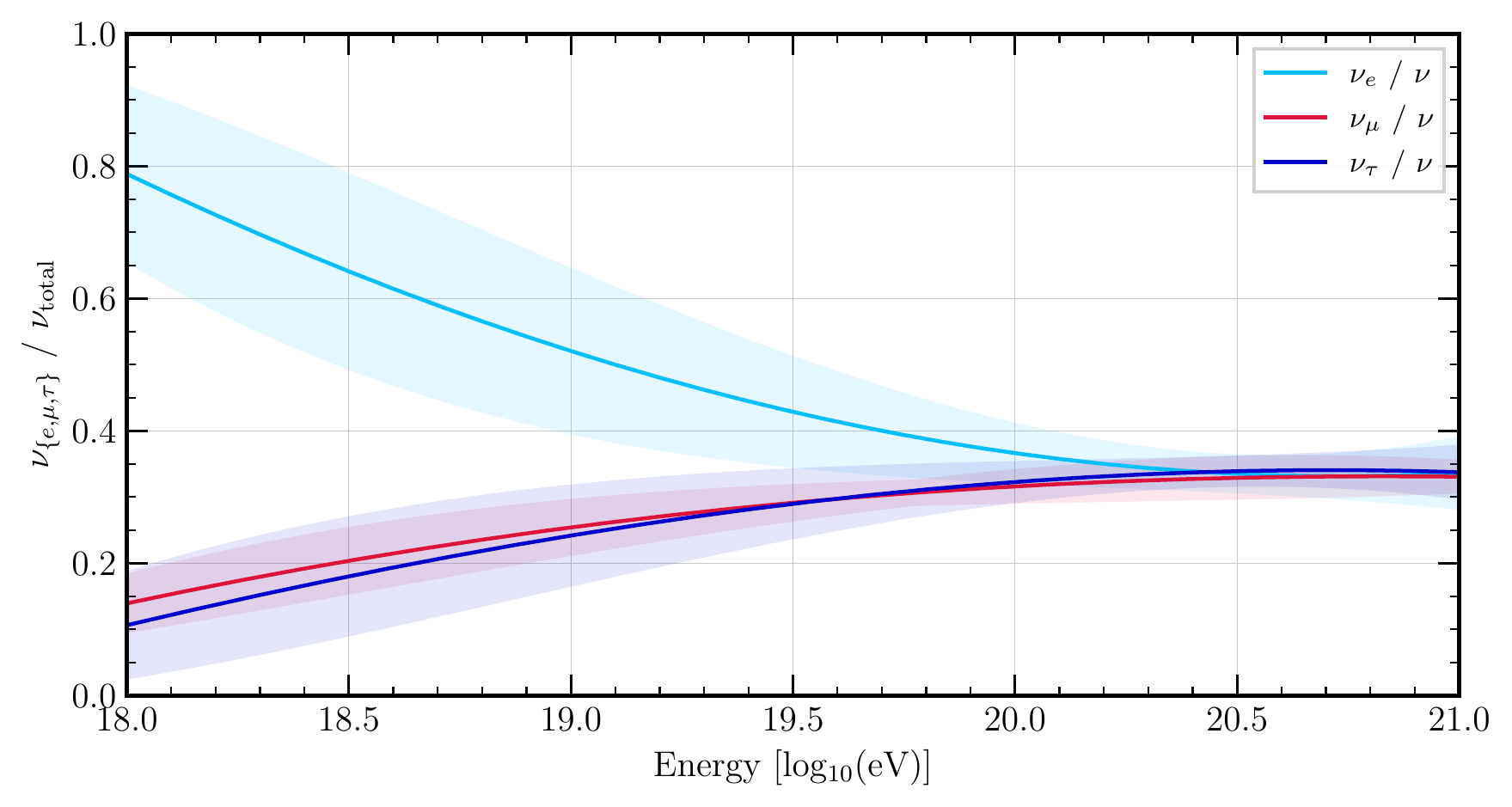}
\caption{The relative fraction of neutrino flavors detected by PUEO via the Askaryan channel.
From 1~EeV to $\sim$10~EeV, PUEO is dominated by electron neutrino events. } 
\label{fig:askaryan_flavor_ratio} 
\end{figure}

As shown in Fig.~\ref{fig:askaryan_flavor_ratio}, at high energies ($>50$~EeV), the Askaryan channel is roughly equally sensitive to all
neutrino flavors. However, at energies below a few EeV, electron neutrinos dominate the Askaryan acceptance. 
The air shower channel, which has comparable sensitivity at several EeV, is almost exclusively sensitive to
tau neutrinos (muon neutrinos have the potential to regenerate through the Earth like $\nu_\tau$ but PUEO's acceptance
to $\nu_\mu$ air showers is several orders of magnitude smaller than $\nu_\tau$).
The Askaryan and EAS channel view similar parts of the sky so any neutrino source
is likely to be detected in both channels if the transient duration is longer than $\sim30$~minutes. 
For these sources, PUEO will be able to measure the $e/\tau$ ratio in a new energy regime, which informs 
not only models of astrophysical neutrino production at the sources, but also tests of
fundamental neutrino physics~\cite{Bustamante:2010nq}. The $e/\tau$ ratio can have strong discriminating power 
for specific neutrino production mechanisms that predict comparable $\mu$ and $\nu_\tau$ fluxes~\cite{palladino19:icecube_flavor}.
 
\section{Conclusion} 

The first detection of UHE neutrinos, whether cosmogenic or astrophysical,
would have profound physics implications for understanding the high-energy
accelerators in our universe. At UHE energies, only neutrinos can
travel the vast cosmological distances providing us acesss to the entirety of the
universe, rather than our local GZK horizon, so only neutrinos can reveal the nature of distant accelerators. The PUEO experiment builds on the
successful ANITA concept to credibly provide the most sensitive instrument for detection
of neutrinos above several EeV. PUEO's sensitivity is derived
from its beamforming trigger, additional antennas, and improved electronics in combination with
the aperture advantages of a high-altitude platform.  PUEO is a versatile instrument that is powerful enough to potentially  
confidently detect UHE neutrinos for the first time for many viable models or, 
alternatively, set the most stringent constraints. 

\acknowledgments 

Development of the PUEO concept is supported by NASA grant 80NSSC20K0775. 
This work was supported by the Kavli Institute for Cosmological
Physics at the University of Chicago and the Center for Cosmology and AstroParticle Physics at The Ohio State University. Computing resources were provided by the
Research Computing Center at the University of Chicago. 

\bibliographystyle{JHEP}
\bibliography{main}

\end{document}